
\input harvmac
\ifx\epsfbox\UnDeFiNeD\message{(NO epsf.tex, FIGURES WILL BE IGNORED)}
\def\figin#1{\vskip2in}
\else\message{(FIGURES WILL BE INCLUDED)}\def\figin#1{#1}\fi
\def\ifig#1#2#3{\xdef#1{fig.~\the\figno}
\goodbreak\midinsert\figin{\centerline{#3}}%
\smallskip\centerline{\vbox{\baselineskip12pt
\advance\hsize by -1truein\noindent\footnotefont{\bf Fig.~\the\figno:} #2}}
\bigskip\endinsert\global\advance\figno by1}
\noblackbox


\def\a{\rightarrow}
\def\aa{\alpha}
\def\b{\beta}
\def\c{\cosh\alpha}
\def\d{\nabla}
\def\l{\lambda}
\def\s{\sinh\alpha}
\def\ss{\sinh^2\alpha}
\def\r{\hat r}
\def\t{\tilde}
\def\Q{{\cal Q}}
\def\R{{\bf R}}
\def\RN{Reissner-Nordstr{\o}m}
\def\({\left(}
\def\){\right)}
\def\frac#1#2{{#1 \over #2}}
\def\[{\left[}
\def\]{\right]}
\gdef\journal#1, #2, #3, 19#4#5{
{\sl #1~}{\bf #2}, #3 (19#4#5)}

\lref\banks{T. Banks, A. Dabholkar, M. Douglas, and M. O'Loughlin,
\journal Phys. Rev., D45, 3607, 1992; T. Banks and M. O'Loughlin,
``Classical and Quantum Production of Cornucopions at Energies Below
$10^{18}$ GeV", Rutgers preprint RU-92-14, hep-th/9206055.}
\lref\aise{ P. Aichelburg and R. Sexl,
\journal Gen. Rel. Grav., 2, 303, 1971.}
\lref\amkl{ D. Amati and C. Klimcik, \journal Phys. Lett., B219, 443, 1989.}
\lref\basf{I. Bars and K. Sfetsos, ``Conformally Exact Metric and Dilaton
in String Theory on Curved Spacetime", USC preprint USC-92/HEP-B2.}
\lref\bekenstein{J. Chase, \journal Commun. Math. Phys., 19, 276, 1970;
J. Bekenstein, \journal Phys. Rev., D5, 1239, 1972.}

\lref\brill{D. Brill, ``Splitting of an Extremal \RN\ Throat via Quantum
Tunneling", \journal Phys. Rev., D46, 1560, 1992.}
\lref\buscher{ T. Buscher, ``Path Integral Derivation of Quantum
Duality in Nonlinear Sigma Models,''
\journal Phys. Lett., B201, 466, 1988 ;
``A Symmetry of the String Background Field Equations,''
\journal Phys. Lett., B194, 59, 1987.}
\lref\btz{M. Banados, C. Teitelboim, and J. Zanelli, ``The Black Hole in
Three Dimensional Spacetime", \journal Phys. Rev. Lett., 69, 1849, 1992,
hep-th/9204099.}

\lref\callan{ C. Callan, R. Myers and M. Perry,
``Black Holes in String Theory", Nucl. Phys. B {\bf 311}, 673 (1988).}
\lref\chs{C. Callan, J. Harvey, and A. Strominger, \journal Nucl. Phys.,
B359, 611, 1991; \journal Nucl. Phys., B367, 60, 1991.}
\lref\cfmp{C. Callan, D. Friedan, E. Martinec and M. Perry, ``Strings in
Background Fields", \journal Nucl. Phys., B262, 593, 1985.}
\lref\csf{E. Cremmer, J. Sherk, and S. Ferrara, \journal Phys. Lett., B74, 61,
1978.}

\lref\unstable{J. McNamara, \journal Proc R. Soc. Lond., A358, 449, 1978;
Y. Gursel, V. Sandberg, I. Novikov, and A. Starobinski, \journal
Phys. Rev. D19, 413, 1979; R. Matzner, N. Zamorano, and V. Sandberg,
\journal Phys. Rev., D19, 2821, 1979; S.~Chandrasekhar and J.~Hartle,
\journal Proc. Roy. Soc. Lond., A384, 301, 1982.}
\lref\pois{E. Poisson and W. Israel, \journal Phys. Rev., D41, 1796, 1990;
A. Ori, \journal Phys. Rev. Lett., 67, 789, 1991; {\bf 68}, 2117 (1992).}
\lref\yurtsever{U. Yurtsever, ``Comments on the Instability of Blackhole
Inner Horizons", Santa Barbara preprint.}

\lref\daha{ A. Dabholkar and J. Harvey, {\it Nonrenormalization of
the Superstring Tension}, {\sl Phys. Rev. Lett.}
{\bf 63} 719 (1989).}
\lref\desa{H. de Vega and N. Sanchez, ``Strings Falling into Spacetime
Singularities", \journal Phys. Rev., D45, 2783, 1992.}

\lref\dghr{ A.~Dabholkar, G.~Gibbons, J.~Harvey and F.~Ruiz,
``Superstrings and Solitons,''
\journal Nucl. Phys., B340, 33, 1990.}

\lref\dvv{ R.~Dijkgraaf, E.~Verlinde and H.~Verlinde,
``String Propagation in a Black Hole Geometry,'' \journal Nucl. Phys., B371,
269, 1992.}
\lref\exactsol{P. Horava, ``Some Exact Solutions of String Theory in Four
and Five Dimensions, \journal Phys. Lett., B278, 101, 1992; D. Gershon,
Exact Solutions of Four Dimensional Black Holes in String Theory",
Tel Aviv preprint TAUP-1937-91}

\lref\ehlers{J. Ehlers, in {\sl Les Theories de la Gravitation}, (CNRS,
Paris, 1959); R. Geroch, ``A Method for Generating Solutions of Einstein's
Equations", \journal J. Math. Phys., 12, 918, 1971.}
\lref\frolov{ V.~Frolov, A.~Zelnikov, and U.~Bleyer,
``Charged Rotating Black Hole from Five Dimensional Point of View,''
\journal Ann. Phys. (Leipzig), 44, 371, 1987.}

\lref\ghs{ D.~Garfinkle, G.~Horowitz and A.~Strominger,
``Charged Black Holes in String Theory,''
\journal Phys.~Rev., D43, 3140, 1991; {\bf D45}, 3888({\bf E}) (1992).}

\lref\garfinkle{D. Garfinkle, private communication.}
\lref\garfbkst{D. Garfinkle, ``Black String Travelling Waves", Oakland U
preprint 92-0403, gr-qc/9209004.}
\lref\geroch{R. Geroch, ``A Method for Generating Solutions to Einstein's
Equation II", \journal J. Math. Phys., 13, 394, 1972.}
\lref\gibbons{ G. Gibbons, ``Antigravitating Black Hole
Solitons with Scalar Hair in N=4 Supergravity", \journal Nucl. Phys., B207,
337, 1982.}
\lref\grla{R. Gregory and R. Laflamme, ``Hypercylindrical Black Holes",
\journal Phys. Rev., D37, 305, 1988; B. Whitt, PhD. Thesis, Cambridge
University, 1988.}

\lref\giwi{
G.~Gibbons and D.~Wiltshire,
``Black Holes in Kaluza-Klein Theory,''
\journal Ann. Phys., 167, 201, 1986;
 {\bf 176}, 393({\bf E}) (1987).}

\lref\gima{ G.~Gibbons and K.~Maeda,
\journal Nucl.~Phys., B298, 741, 1988.}
\lref\gist{S. Giddings and A. Strominger, ``Dynamics of Extremal Black Holes",
\journal Phys. Rev., D46, 627, 1992.}
\lref\gistexact{S. Giddings and A. Strominger, ``Exact Black Five Branes in
Critical Superstring Theory", \journal Phys. Rev. Lett., 67, 2930, 1991.}

\lref\giveon{ A.~Giveon, ``Target Space Duality and Stringy Black Holes,''
\journal Mod. Phys. Lett., A6, 2843, 1991;
 E.~Kiritsis, ``Duality in Gauged WZW Models,'' \journal Mod. Phys. Lett.,
 A6, 2871, 1991; P. Ginsparg and F. Quevedo, ``Strings on Curved Spacetime:
 Black Holes Torsion and Duality", Los Alamos preprint LA-UR-92-640,
 hep-th/9202092.}
\lref\grha{R. Gregory and J. Harvey, ``Black Holes with a Massive Dilaton",
Enrico Fermi Preprint EFI-92-49, hepth/9209070.}

\lref\guven{ R.~Guven, \journal Phys.  Lett., B191, 275, 1987.}

\lref\hawking{ S. Hawking, ``Gravitational Radiation From Colliding
Black Holes," \journal Phys. Rev. Lett., 26, 1334, 1971.}

\lref\hawkevap{S. Hawking, ``Particle Creation from Black Holes",
{\sl Commun. Math. Phys.}, {\bf 43}, 199, (1975).}
\lref\hael{S. Hawking and Ellis, {\sl The Large Scale Structure of Spacetime,}
Cambridge University Press, 1973.}
\lref\hast{J. Harvey and A. Strominger, ``Quantum Aspects of Black Holes",
to appear in the proceedings of the 1992 Trieste Spring School on String
Theory and Quantum Gravity, hep-th/9209055.}

\lref\howi{ C.~Holzhey and F.~Wilczek,
``Black Holes as Elementary Particles,'' \journal Nucl. Phys., B380, 447,
1992.}

\lref\hoho{ J.~Horne and G.~Horowitz,
``Exact Black String Solutions in Three Dimensions,''
\journal Nucl. Phys., B368, 444, 1992.}
\lref\hohorot{J.~Horne and G.~Horowitz, ``Rotating Dilaton Black Holes",
\journal Phys. Rev., D46, 1340, 1992.}
\lref\hohomass{J.~Horne and G.~Horowitz, ``Black Holes Coupled to a Massive
Dilaton", Santa Barbara preprint UCSBTH-92-17, hep-th/9210012.}

\lref\hhs{ J.~Horne, G.~Horowitz, and A.~Steif,
``An Equivalence Between Momentum and Charge in String Theory,''
\journal Phys. Rev. Lett., 68, 568, 1992.}

\lref\host{ G.~Horowitz and A.~Strominger, ``Black Strings
and $p$-Branes,''
\journal Nucl. Phys., B360, 197, 1991.}
\lref\jjp{I. Jack, D. Jones and J. Panvel, ``Exact Bosonic and Supersymmetric
String Black Hole Solutions", Liverpool preprint, LTH 277 (1992)
hep-th/920139.}
\lref\klopp{R. Kallosh, A. Linde, T. Ortin, A. Peet, and A. Van Proeyen,
``Supersymmetry as a Cosmic Censor", Stanford preprint SU-ITP-92-13.}
\lref\kinnersley{W. Kinnersley, ``Generation of Stationary
Einstein-Maxwell Fields",
\journal J. Math. Phys., 14, 651, 1973.}

\lref\kiya{ K.~Kikkawa and M.~Yamasaki,
\journal Phys. Lett., B149, 357, 1984 ;
N.~Sakai and I.~Senda,
\journal Prog. Theor. Phys., 75, 692, 1986 ;
V.~Nair, A.~Shapere, A.~Strominger and F.~Wilczek,
\journal Nucl. Phys., B287, 402, 1987.}

\lref\leutwyler{ H.~Leutwyler,
\journal Arch.~Sci., 13, 549, 1960;
P.~Dobiasch and D.~Maison,
\journal Gen. Rel. Grav., 14, 231, 1982 ;
A.~Chodos and S.~Detweiler,
\journal Gen. Rel. Grav., 14, 879, 1982; L. Simon, \journal Gen. Rel. Grav.,
17,
439, 1985.}
\lref\ils{N. Ishibashi, M. Li and A. Steif, ``Two Dimensional Charged Black
Holes in String Theory", \journal Phys. Rev. Lett.,
67, 3336, 1991; M. McGuigan, C. Nappi and S. Yost, ``Charged Black Holes in Two
Dimensional String Theory", \journal Nucl. Phys., B375, 421, 1992.}
\lref\msw{G. Mandel, A. Sengupta and S. Wadia, \journal Mod. Phys. Lett.,
A6, 1685, 1991.}

\lref\myers{ R. C. Myers,  ``Superstring Gravity and Black Holes",
Nucl. Phys. {\bf B289} (1987) 701.}

\lref\mype{ R.~Myers and M.~Perry,
``Black Holes in Higher Dimensional Space-times,''
\journal Ann. Phys., 172, 304, 1986.}
\lref\ortin{T. Ortin, ``Electric-Magnetic Duality and Supersymmetry in
Stringy Black Holes", Stanford preprint SU-ITP-92-24, hepth/9208078.}

\lref\sipe{ M.~Simpson and R.~Penrose,
\journal Int. J. Theor. Phys., 7, 183, 1973.}

\lref\psstw{ J.~Preskill, P.~Schwarz, A.~Shapere, S.~Trivedi,
and F.~Wilczek,
``Limitations on the Statistical Description of Black Holes,''
\journal Mod. Phys. Lett., A6, 2353, 1991.}

\lref\rove{ M.~Ro\v cek and E.~Verlinde,
``Duality, Quotients, and Currents,'' \journal Nucl. Phys., B373, 630, 1992
hep-th/9110053;
A. Giveon and M. Ro\v cek, ``Generalized Duality in Curved
String Backgrounds", \journal Nucl. Phys., B380, 128, 1992.}

\lref\sen{ A.~Sen, ``$O(d) \otimes O(d)$ Symmetry of the Space of
Cosmological Solutions in String Theory, Scale Factor Duality,
and Two Dimensional Black Holes,'' \journal Phys. Lett., B271, 295, 1991;
A.~Sen, ``Twisted Black $p$-brane Solutions in String
Theory,'' \journal Phys. Lett., B274, 34, 1992;
S.~Hassan and A.~Sen,
``Twisting Classical Solutions in Heterotic String Theory,''
\journal Nucl. Phys., B375, 103, 1992}

\lref\meve{K.~Meissner and G.~Veneziano,
``Symmetries of Cosmological Superstring
Vacua,'' \journal Phys. Lett., B267, 33, 1991;
``Manifestly $O(d,d)$ Invariant Approach
to Space-Time Dependent String Vacua,'' \journal Mod. Phys. Lett., A6, 3397,
1991 hep-th/9110004.}
\lref\masc{J. Maharana and J. Schwarz, ``Noncompact Symmetries in String
Theory, Caltech preprint CALT-68-1790.}

\lref\senrev{A. Sen, `` Black Holes and Solitons in String Theory",
Tata preprint TIFR-TH-92-57, hep-th/9210050.}
\lref\senrot{A. Sen, ``Rotating Charged Black Hole in String Theory",
\journal Phys. Rev. Lett., 69, 1006, 1992 hep-th/9204046.}
\lref\senem{A. Sen, ``Electric Magnetic Duality in String Theory",
Tata preprint TIFR-TH-92-41, hep-th/9207053.}
\lref\senemst{A. Sen, ``Macroscopic Charged Heterotic String", Tata preprint
TIFR-TH-92-29, hep-th/9206016.}
\lref\sfetsos{K. Sfetsos, USC preprint USC-92/HEP-S1.}
\lref\stw{A. Shapere, S. Trivedi, and F. Wilczek, ``Dual Dilaton Dyons",
\journal Mod. Phys. Lett.,
A6, 2677, 1991.}

\lref\smpo{ E.~Smith and J.~Polchinski,
``Duality Survives Time Dependence,'' \journal Phys. Lett., D263, 59, 1991;
A.~Tseytlin, ``Duality and Dilaton,''
\journal Mod. Phys. Lett., A6, 1721, 1991.}

\lref\steif{ G.~Horowitz and A.~Steif,
\journal Phys.~Rev.~Lett., 64, 260, 1990;
\journal Phys. Rev., D42, 1950, 1990.}

\lref\wald{ R.~Wald,
{\it General Relativity} (U.~of Chicago Press, Chicago), 1984.}

\lref\witten{ E.~Witten, ``On String Theory and Black Holes,''
\journal Phys. Rev., D44, 314, 1991.}

\Title{\vbox{\baselineskip12pt\hbox{UCSBTH-92-32}
\hbox{hep-th/9210119}}}
{\vbox{\centerline {The Dark Side of String Theory:}
       \centerline {Black Holes and Black Strings}}}
\centerline{{ Gary T. Horowitz}\footnote{$^*$}
{To appear in the proceedings of the 1992
Trieste Spring School on String Theory
and Quantum Gravity}}
\vskip.1in
\centerline{\sl Department of Physics}
\centerline{\sl University of California}
\centerline{\sl Santa Barbara, CA 93106-9530}
\centerline{\sl gary@cosmic.physics.ucsb.edu}
\bigskip
\centerline{\bf Abstract}
Solutions to low energy string theory describing black holes and
black strings are reviewed.
Many of these solutions can be obtained by applying
simple solution generating transformations to the Schwarzschild metric.
In a few cases, the corresponding exact conformal field theory
is known.
Various properties of these solutions are discussed
including
their global structure, singularities, and  Hawking temperature.

\Date{10/92}

\newsec{INTRODUCTION}

One of the most intriguing predictions of general relativity is the
existence of black holes.  There is now good observational evidence that
black holes exist throughout the universe on scales from a solar mass
(which are seen in binary star systems) up to millions of solar masses
(which are seen in the center of galaxies and quasars). For these black
holes, general relativity provides an adequate description at this time.
However, it has been suggested that much smaller black holes could have
been formed in the early universe. These black holes will become even smaller
through the emission of
Hawking radiation.  Even the large black holes we see
today will evaporate in the distant future if the temperature of the
cosmic background radiation becomes less than their Hawking temperature.

In the late stages of this evaporation, general relativity is expected
to break down and be replaced by a quantum theory of gravity.  Since
string theory is a promising candidate for a consistent quantum theory
of gravity, it is of interest to examine black holes in string
theory. As a first step, one should study the classical
black holes solutions in this theory. This is what I plan to do in
these lectures.   We will concentrate on black holes with electric or
magnetic charge. For these holes, the predictions of string theory
differ from those of general relativity long before Planck scale
curvatures are reached. The reason for this difference is the presence of
a scalar field called the dilaton. We will see that the dilaton dramatically
changes the  properties of extremal black holes. For example, the extremal
solutions can be completely free of curvature singularities. They can also
repel each other. In addition to their
possible astrophysical interest, charged black holes provide an ideal
setting for studying the late stages of Hawking evaporation. The modifications
predicted by string theory may help to resolve some of the puzzles associated
with this process \banks.

In addition to black holes, it turns out
that string theory has solutions  describing  one-dimensional extended
objects surrounded by event horizons i.e. black strings.
We will see that these solutions can have unusual causal structure, and
provide some insight into the properties of singularities in string theory.
Most importantly, they are closely connected to fundamental strings themselves.
A black string carries a charge per unit length, and in the extremal limit,
the solution reduces
to precisely the field outside a  straight fundamental string.
(There are  other extended black hole solutions in string theory corresponding
to black membranes or black $p$-branes \host\ but I will not discuss them
here.)

Let me clarify what I mean by classical solutions to string theory
by distinguishing
three different levels of approximation.
In increasing importance (and difficulty) we have

1) Perturbative solutions of the low energy classical action

2) Exact solutions of the low energy classical action

3) Exact solutions of the full classical action

Solutions of the first type are obtained by considering (tree level)
string scattering in flat spacetime.
This was historically the first class of solutions discussed,
but they are not very useful for describing black holes.
Solutions of the third type are believed to correspond to two dimensional
conformal field theories. These are ultimately what we after, but at this time,
there are only a few black hole and black string
solutions of this type known. Since the  known solutions exist in unphysically
low spacetime dimensions,
I will focus mostly on solutions of the second type.
These should be good approximations to  exact solutions whenever
the curvature is small compared to the Planck curvature. This can include the
horizons as well as the region outside the black hole,
but not of course a neighborhood of the singularity.
It turns out that
many of the solutions of the second type
can be obtained by using solution generating
techniques. These are similar to the transformations  which have been found
for the vacuum Einstein \ehlers, and Einstein-Maxwell \kinnersley\ equations.
Although the emphasis will be on the classical properties of these solutions,
some basic aspects of Hawking evaporation such as the Hawking temperature
will be discussed.

In Sec. 2, we will first review the standard black hole solutions in the
Einstein-Maxwell theory. We then describe the low energy string field
equations and discuss two
methods of generating solutions to these equations. In Sec. 3 we
begin our investigation of black holes in string theory by discussing the
string analog of the \RN\ solution.
We describe several generalizations of this solution in Sec. 4. These include
black holes with rotation,  a (physically
expected) mass term for the dilaton, and in other spacetime dimensions.
The black string solutions are discussed in Sec. 5. Finally,
in Sec. 6, we
briefly consider the subject of singularities in exact solutions to the
full classical string action, and conclude with
some open problems.

In reviewing a subject of this type where there are a large number of known
solutions,
one faces a decision about how much information to include.
My aim is to make this review
self contained and useful as a reference, while keeping it clear and
readable. Thus, I have not included the most general black hole solution
known at this time. Instead, I describe the basic charged black hole solution
in Sec. 3 and then discuss how it is modified when one includes
rotation, other dimensions, etc. in Sec. 4. Similarly, in Sec. 2, I do not
include the most general solution generating transformation, but
restrict attention to the two which are most useful in obtaining the
black hole and black string solutions. The reader interested in
pursuing a topic further can use the references as a guide to the
literature.  For another recent review of this subject, see \senrev.

\newsec{PRELIMINARIES}

Before discussing the actual black holes solutions it is necessary to
cover a few preliminaries. First we review the standard black hole solutions
in Einstein-Maxwell theory, and the calculation of their Hawking temperature.
Then we discuss  the low energy string
action and associated field equations. Finally, we describe two
methods of generating solutions to these equations. We will see that many
of the solutions of interest can
be found by  applying these methods to the Schwarzschild solution.

\subsec{Black Holes in Einstein-Maxwell Theory}

\ifig\tschwarz{The Penrose diagram for the
maximally extended Schwarzschild solution.} {\epsfbox{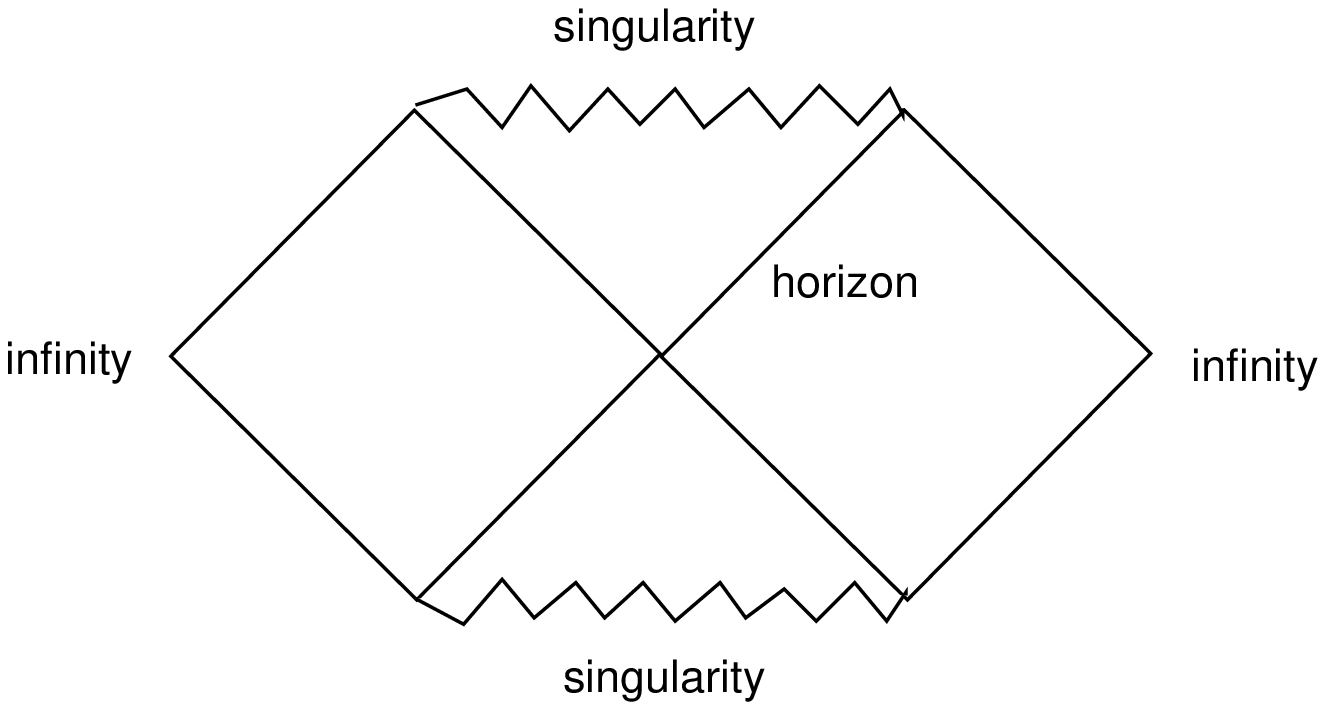}}

As is well known, uncharged static black holes are
described by the Schwarzschild solution.  This solution takes the form
\eqn\schwarz{
ds^2 = - \(1-{2M\over r}\) dt^2 + \(1-{2M \over r}\)^{-1} dr^2 + r^2 d\Omega}
where $M$ is the mass of the black hole. The
global structure of  Schwarzschild is conveniently described by a
Penrose diagram in which light rays move along 45$^\circ$ lines  and
infinity has been brought to a finite distance by a conformal rescaling\foot{
For a more detailed discussion of Penrose diagrams see~\hael\ or the
lectures in this volume by Harvey and Strominger~\hast.}.
This is shown in \tschwarz. (One can interpret this two dimensional figure
either as representing the $r-t$ plane, or the entire spacetime where each
point represents a two sphere of
spherical symmetry.)
The event horizon is located at $r=2M$ where $g_{tt} = 0$.  Since Schwarzschild
is time reversal invariant, the maximally extended spacetime contains a white
hole as well as a black hole, and a second
asymptotically flat region.

\ifig\trnbh{The Penrose diagram for the \RN\ solution with $Q<M$.}
{{\epsfysize = 5in \epsfbox{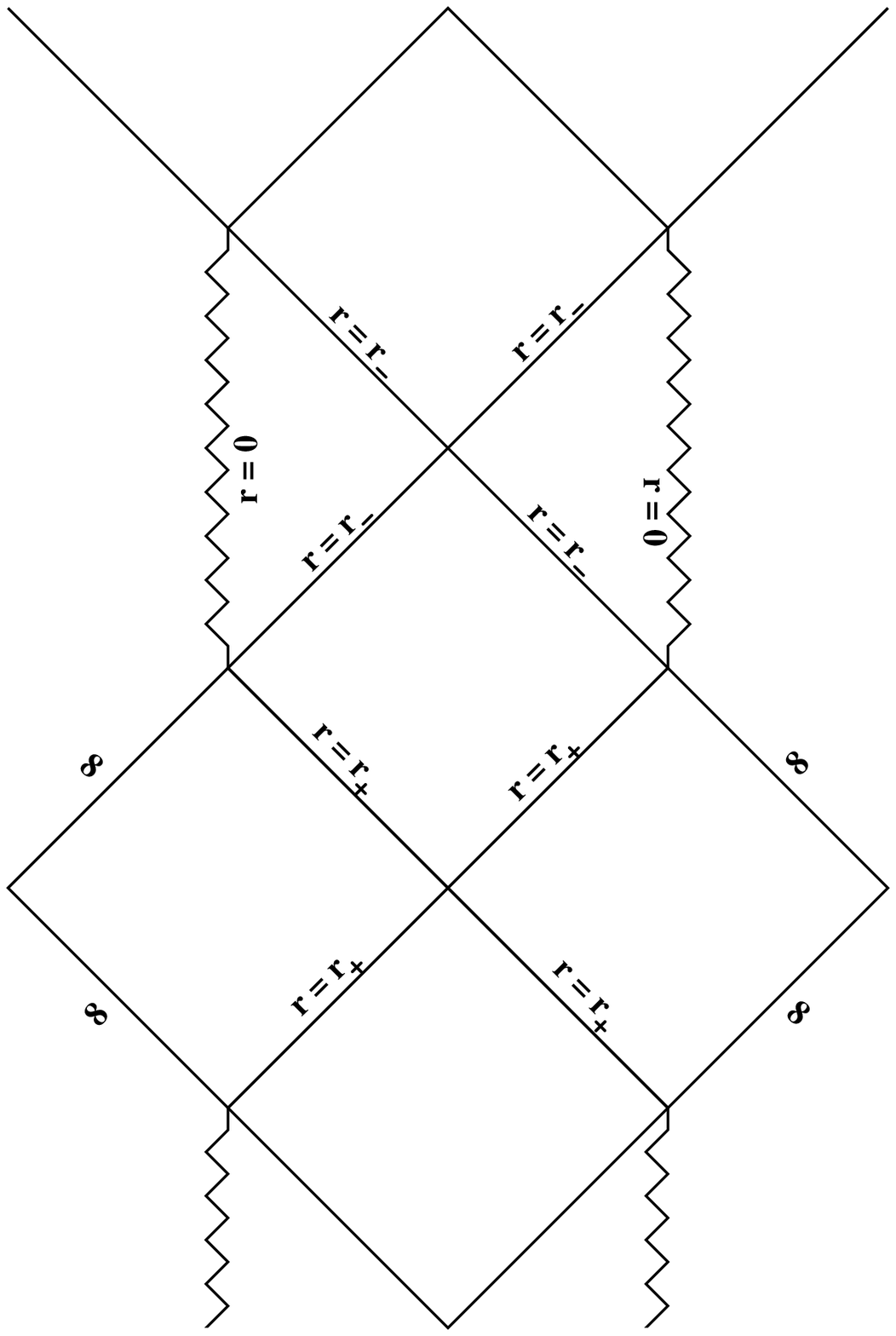}}}

A charged black hole in general relativity is described by the
\RN\ solution which has the metric
\eqn\rn{
ds^2 = - \(1-\frac{2M}{r} + \frac{Q^2}{r^2}\) dt^2 + \(1-\frac{2M}{r} +
      \frac{Q^2}{r^2}\)^{-1} dr^2
      + r^2 d\Omega}
together with a  Maxwell field given by $F_{rt} = Q/r^2$
for an electrically charged  hole and
$F_{\theta\varphi} = Q\sin\theta$ for a magnetically charged hole.  Its global
structure is quite different from Schwarzschild and depends on the relative
size of $Q$ and $M$.
(We are using  geometrical units $G=c=1$
in which the charge on an electron is equivalent to $10^{-6}$ gm. Since this
is much larger than the mass of an electron, it is relatively easy
-- in principle -- to create a black hole with  $Q$ of order $M$.)
For $0<|Q|<M$ there are now
two zeros of $g_{tt}$ at $r = r_\pm$ where
\eqn\rpm{
r_{\pm} \equiv M \pm \sqrt{M^2-Q^2}}
which correspond to two horizons.
There is an event horizon at
$r=r_+$ and an inner horizon at $r=r_-$.
The Penrose diagram is shown in \trnbh.
The significance of the inner horizon is  the following. Starting with
initial data on an asymptotically flat spacelike surface, one is guaranteed a
unique evolution only up to the inner horizon. After that, the evolution will
be affected by boundary conditions at the singularity.
Note that the singularity is now timelike, so an
observer falling in is not forced to hit it.  He can continue into
another asymptotically flat region of spacetime. (The maximally extended
spacetime contains an infinite number of such regions.)
In fact, he must literally
work hard to reach the singularity since freely falling observers
avoid it: \RN\ is timelike geodesically complete. In light of this, you might
be tempted to carry a few charges with you in case you fall into a black
hole. Until recently, it was widely believed that this would not help.
The inner horizon was known to be unstable~\unstable\
and slightly nonspherical charged
collapse was thought to result in a spacetime resembling Schwarzschild.
However recent work~\pois\ has indicated that the singularity at the inner
horizon
might be much more mild. (Although this is not yet settled~\yurtsever.)
 In any case, a journey  through the inner horizon would not be uneventful.
Immediately after crossing the inner horizon, one would be face to face with
a curvature singularity!

\ifig\trnext{The  Penrose diagram for the \RN\ solution with $Q=M$. A $t=$
const
surface does not actually hit the singularity, but contains an infinite
throat.}{{\epsfysize = 5in \epsfbox{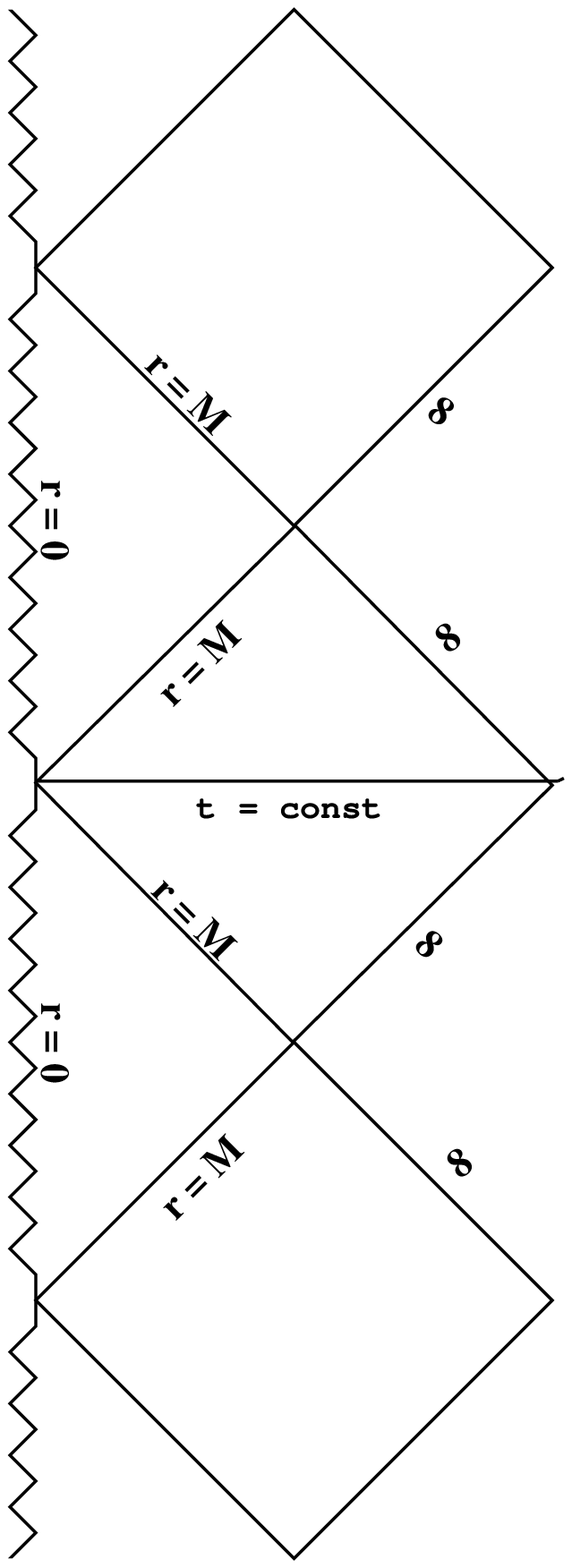}}}

 For $Q=M$,
the horizons
coincide: $r_+ = r_- = M$.  The Penrose diagram is now shown in \trnext.
A $t=$ const surface now looks
like it hits the singularity, but this is just an artifact of the
conformal rescaling.  With $Q=M$, the proper distance to the horizon
$r=M$ from a point $r_0 > M$ along a $t=$const, radial curve is
\eqn\length{
L= \int^{r_0}_M \frac{dr}{\left(1-\frac{M}{r}\right)} = \infty}
So a $t=$const surface asymptotically resembles a cylinder as $r \a M$.
The geometry describes an infinite throat.
In a sense, the horizon and other asymptotic flat region (which exist for
$Q<M$) has
been pushed off to infinity.  However, even though the horizon is
infinitely far away in  spacelike directions, it is only a {\it finite}
distance away in timelike or null directions.  Observers can still fall
into the black hole in a finite proper time.

\ifig\tnaksing{The Penrose diagram for the \RN\ solution with $Q>M$.}
{\epsfbox{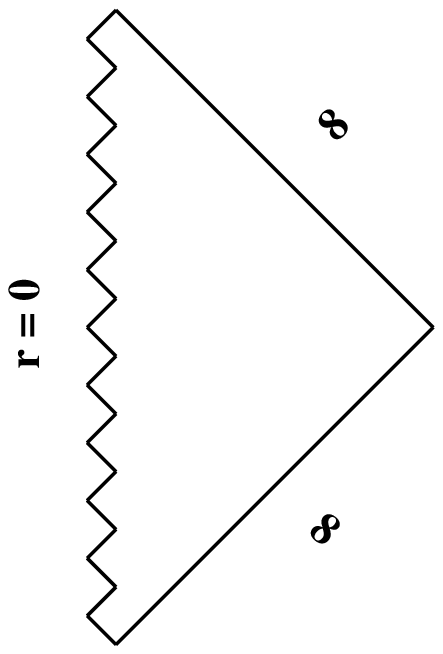}}

Finally, for $Q>M$, the \RN\ solution
does not describe a black hole at
all but rather a naked singularity. The Penrose diagram is shown in
\tnaksing. For this reason, the $Q=M$
solution is called the extremal black hole.   It has the largest
possible ratio of charge to mass.

The Hawking temperature of a  static
black hole can be calculated in several ways.
Hawking's original calculation \hawkevap\
involved studying quantum matter fields in
the black hole background. It was latter realized that
one could compute this temperature by simply analytically continuing in $t$
and requiring that the resulting Riemanian space be nonsingular. This
requires a  periodic identification in imaginary time, and the temperature is
one over this period. Physically, this instanton is related to
a black
hole in thermal equilibrium with a gas, in the approximation where the
energy density of the gas is neglected. It will be convenient later to
have a simple formula for the Hawking temperature of a general static,
spherically symmetric black hole
solution. Suppose the $r-t$ plane
has metric
\eqn\caltemp{ds^2 = -\lambda dt^2 + dr^2/f}
If there is an event horizon
at $r=r_0$, then near this horizon, $\lambda \approx \l'(r_0) \xi$ and
$f\approx f'(r_0)\xi$ where
$\xi = r-r_0$. (We are assuming here that the event
horizon is not degenerate.) We now set $\tau = i t$ and
$\rho = 2\sqrt{\xi/f'(r_0)}$.
The resulting metric is
\eqn\caltem{  d\rho^2 + {\l'(r_0) f'(r_0)\over 4} \rho^2 d\tau^2}
To avoid a conical singularity at $\rho =0$ we must identify $\tau $ with
period $4\pi/ \sqrt{\l' f'}$. Thus the Hawking temperature is
\eqn\gentemp{T_H = {\sqrt{\l'(r_0) f'(r_0)}\over 4\pi}}

Applying this to the \RN\ metric yields
\eqn\rntemp{
T_H=\frac{\sqrt{\left(M^2-Q^2\right)}}{2\pi\left(M+\sqrt{M^2-Q^2}\right)^2}}
A  charged
black hole will preferentially radiate away its charge. However  the amount
of charge that is actually radiated away depends on the charge to
mass ratio of the particles
in the theory. If this ratio is sufficiently small, most of the radiation
will be in the form of neutral particles and $Q$ will be essentially constant.
This is likely to be true for magnetically charged black holes.
In this case, the black hole will
evolve toward its extremal limit. The Hawking temperature~\rntemp\
vanishes as $Q\rightarrow M$.  This suggests that extremal
charged black holes may be quantum mechanically stable. (There
is still a possibility that extremal quantum black holes can
bifurcate \brill.)
This is  consistent
with ideas of cosmic censorship. Although cosmic censorship is usually
discussed in the context of classical general relativity, it is reassuring
that as you approach the extremal limit,
the Hawking radiation turns off.  One does not continue to radiate
to a naked singularity.

\subsec{The Equations of Motion}

We will work with part of
the low energy action to heterotic  string theory. The most general situation
we will consider is described by a metric $g_{\mu\nu}$, a dilaton $\phi$,
a Maxwell field $F_{\mu\nu}$, and a three form $H_{\mu\nu\rho}$. The Maxwell
field is associated with a $U(1)$ subgroup of the gauge group.
We will set the rest of
the gauge
field to zero as well as the fermions.
The three form $H_{\mu\nu\rho}$ is  related to
a two-form potential  $B_{\mu\nu}$  and the gauge field $A_\mu$
by $H = dB - A \wedge F$
so that $dH = - F\wedge F $. (Since we will keep only terms with two
derivatives
or less, it is consistent to drop the Lorentz Chern Simons term which also
appears
in the definition of $H$.)

These fields are governed by the following action\foot{Our conventions for
the curvature follow those of Wald \wald.} \cfmp:
\eqn\action{
S= \int d^D x \sqrt{-g}\ e^{-2\phi} \Bigl[ \Lambda +R+4(\nabla \phi)^2 -
F_{\mu\nu}F^{\mu\nu} - \frac{1}{12} H_{\mu\nu\rho}
H^{\mu\nu\rho}\Bigr]}
where $\Lambda$ is a constant which is related to the spacetime dimension
$D$ and the central charge of a possible ``internal" conformal field theory.
(Since this internal part of the solution only affects black holes
by changing the value of $\Lambda$, we will not discuss it further.)
It is clear from this action that $e^\phi$
plays the role of a coupling constant. It governs the strength of quantum
corrections.
The complete string action includes higher order corrections $R^2, R^3, F^4$
etc. These can be neglected when discussing black hole solutions
provided the size of the hole is much larger than the Planck length. In
this case,
the higher order corrections will induce, at most, small changes in the
solution until one is well inside the event horizon and near the singularity.
(The effect of  some of these higher order corrections
on the Schwarzschild solution has been discussed in~\callan\myers.)
It is, of course, important to
find exact black hole solutions, but at the present time this is known
only in  two spacetime dimensions~\witten.

The equations of motion which follow from this action are
\eqna\fdeq
$$\eqalignno{&R_{\mu\nu} + 2\d_\mu\d_\nu \phi - 2F_{\mu\lambda} {F_\nu}^\lambda
      - {1 \over 4} H_{\mu\lambda\sigma}
       {H_\nu}^{\lambda\sigma} =0   &\fdeq a \cr
 &\d^\nu (e^{-2\phi} F_{\mu\nu}) + {1 \over 12}
      e^{-2\phi}H_{\mu\nu\rho} F^{\nu\rho}
	    = 0  & \fdeq b \cr
 &\d^\mu (e^{-2\phi} H_{\mu\nu\rho}) = 0 & \fdeq c \cr
&4\d^2\phi -4(\d\phi)^2 + \Lambda + R - F^2 -{1\over 12} H^2 = 0 &\fdeq d\cr}$$
The dilaton appearing in these equations is massless. But it is expected that
when supersymmetry is broken, the dilaton will acquire a mass. We will
consider black hole solutions with a massive dilaton in Sec. 4.5. It turns out
that black
holes which are small compared to the Compton wavelength of the dilaton will
resemble the massless dilaton solutions. So it is worthwhile to begin by
studying the black hole solutions to ~\fdeq{}\ without the mass term.

\subsec{Generating Solutions}

At first sight it appears difficult to find exact solutions to \fdeq{}.
The presence of the
exponential of the dilaton makes the field equations rather complicated.
However it turns out that if one considers spacetimes with a symmetry,
one can generate new solutions from old ones by a simple
transformation~\sen\meve\masc.
For a solution with several symmetry directions
there is a large class of new solutions,
but here we will consider just the simplest case of a single
symmetry.
To illustrate the type of transformation we will use, let us first consider
Kaluza-Klein theory. The five dimensional vacuum Einstein action, when
restricted to spacetimes that are independent of one direction $x$, is
equivalent to the following action for a four dimensional metric,
Maxwell field, and scalar:
\eqn\kkaction{ S=\int d^4x \sqrt{-g} \left(R -2 (\d \Phi )^2 -
	    e^{-2\sqrt 3 \Phi} F^2
				      \right) }
The components $g_{\mu 5}$ of the five dimensional metric are essentially
the four dimensional vector potential. This theory also appears difficult
to solve exactly. However, once the connection with five dimensional general
relativity is understood, it is easy to generate nonvacuum solutions
starting from a static vacuum solution. Given a static
four dimensional vacuum metric, one
can take its product with $R$ to obtain a five dimensional solution
with two symmetry directions.
One can now boost this solution in the fifth direction. This clearly
still satisfies the five dimensional field equations. However when
reinterpreted
in four dimensions, one obtains a solution with nonzero Maxwell field and
dilaton. In particular, one can find charged black hole solutions to~\kkaction\
starting from Schwarzschild this way~\leutwyler\giwi.

In heterotic string theory, the situation is slightly different. The extra
spacetime
coordinates are divided into left-moving and right-moving parts.
Only half of these
are added to the theory. This results in the  low energy
gauge fields. Nevertheless
there is still a way to ``boost" a  static
uncharged solution to obtain a charged one. To be explicit, we
start with any
static solution ($g_{\mu\nu}, B_{\mu\nu},
\phi$) to \fdeq{}\ with $A_\mu=0$. Since the solution is static, rather than
just
stationary,
$g_{ti}
=0$. (For simplicity, we will further assume $B_{ti}=0$.)
Then one can obtain a  one parameter family of
solutions with $B_{\mu\nu}$ unchanged,
$g_{ij}$ unchanged and \sen
\eqn\trans{\eqalign{\t g_{tt} = &{g_{tt} \over [1+ (1+g_{tt})\sinh^2\aa]^2}\cr
    \tilde A_t =& -{(1+g_{tt})\sinh 2\aa
	\over  2\sqrt 2 [1+ (1+g_{tt})\sinh^2\aa]} \cr
    e^{-2\tilde \phi} = & e^{-2\phi} [1+ (1+g_{tt})\sinh^2\aa] \cr}}
where $\aa$ is an arbitrary parameter. When $\aa=0$, the transformation reduces
to the identity. This formula can be generalized
to include nonzero $g_{ti}$ and $B_{ti}$. There is an analogous transformation
for spacelike symmetries.

There is a second transformation which will play an important role in
our discussion of black strings. This is a discrete transformation
which relates solutions of \fdeq{}\ with a symmetry\foot{The two
transformations
that we describe here are actually part of an $O(2,1)$ symmetry group \sen.}.
For simplicity, we again
set $A_\mu=0$ and let ($g_{\mu\nu}, B_{\mu\nu},
\phi$) be a solution to \fdeq{}\ which is independent of $x$.
Then ($\tilde g_{\mu\nu}, \tilde B_{\mu\nu},\tilde \phi$)
is also a solution where \buscher
\eqn\sigmadual{\eqalign{
 \tilde g_{xx} & = 1/g_{xx}, \qquad  \t g_{x\aa} = B_{x\aa}/ g_{xx} \cr
  \t g_{\aa\b} & = g_{\aa\b} - (g_{x\aa}g_{x\b} - B_{x\aa}B_{x\b})/g_{xx} \cr
  \t B_{x\aa} & = g_{x\aa}/g_{xx}, \qquad
		\t B_{\aa\b} = B_{\aa\b} -2 g_{x[\aa} B_{\b]x}/g_{xx} \cr
  \tilde \phi & = \phi - {1\over 2} \log g_{xx} \cr }}
and $\aa,\b$ run over all directions except $x$. This transformation is
sometimes called spacetime (or target space) duality. If $x$ is periodic,
\sigmadual\ is not just another solution to the field equations.
It is known that string theory has the remarkable property that different
spacetime geometries can correspond to the same conformal field theory
\kiya\smpo. The transformation \sigmadual\ is
a generalization
of the $r \a 1/r$ symmetry of strings moving on a circle of radius $r$.
As in that flat spacetime example, one can show \rove\ that if $x$
is compact, the tilded solution \sigmadual\
is physically equivalent to the original solution.
 There are other examples of duality for solutions without a
symmetry direction.  It is conceivable that every solution has at least
one dual description.

We now describe a general result on the effect of spacetime duality
on asymptotically defined conserved quantities, and show how it can
be applied to
black strings.
Let $g_{\mu\nu}$, $B_{\mu\nu}$ and $\phi$ be a solution to the low energy
field equations which is independent
of $x$, and is asymptotically flat in the transverse direction. Then one
can define the mass per unit length, or more generally the
ADM energy momentum per unit length $P_\mu$. We can also define
a  charge per unit length associated with the antisymmetric tensor field by
$\Q = \int e^{-2\phi}\ {}^*H/V_{D-3}$ where $V_{D-3}$ is the volume of a unit
$D-3$ sphere and the integral is over
the $D-3$ sphere at
fixed time, fixed $x$, and large transverse distance.
Since the dual
solution \sigmadual\
is also translationally invariant and asymptotically flat one can
define an energy momentum and charge per unit length associated with it. One
can now ask what is the relation between these quantities. One finds the
surprising result \hhs
\eqn\pqduality{\tilde \Q = P_x, \qquad \tilde P_x = \Q, \qquad \tilde P_\alpha
=P_\alpha}
In other words, the effect of duality is simply to interchange the
charge and the momentum in the symmetry direction. Since these solutions
represent the same conformal field theory, one learns that the charge
associated
with $H$ is equivalent to momentum in string theory.

One can use this to add charge to any solution which is both static and
translationally invariant as follows. One first boosts the solution to obtain
$P_x \ne 0$ and then applies duality to convert this momentum into charge.
This result may have applications independent of black strings, but
we use it in Sec. 5 to obtain black string solutions. Since this charge is
equivalent to momentum, why bother constructing the charged solutions?
The answer is that having an alternative description of the solution is very
useful for making contact with other results in string theory. It also
illustrates what properties of spacetime are well defined in string theory
(i.e. duality invariant) and which are not.

The result \pqduality\ is somewhat reminiscent of Kaluza-Klein theory,
where spacetime
momentum in an internal direction gives rise to charge in the lower dimensional
space. But there is a crucial difference. In the present case, the charge
arises in the higher dimensional space and is associated with a separate field.
It is not part of the  higher dimensional metric.

\newsec{STRING ANALOG OF REISSNER-NORDSTR{\O}M }

We can now begin our discussion of
black hole solutions to the low energy string equations \fdeq{}.
In this section
we will consider
the most physical case of four spacetime dimensions. The appropriate
boundary conditions are that
the spacetime  be asymptotically flat and the dilaton  approach a constant
at infinity
which we will take to be zero. (We will see how to obtain other asymptotic
values of the
dilaton in Sec. 4.1.) These boundary conditions require $\Lambda =0$. For
simplicity, in this section we will also set $H=0$.
The metric in \action\ is the natural one to use since
it is the one that strings directly couple to.  But in order to compare
with  general relativity, it is convenient to rescale $g_{\mu\nu}$ by
$e^{-2\phi}$ to get a metric with the standard Einstein action: $g^E_{\mu\nu} =
e^{-2\phi} g_{\mu\nu}$.  The action now becomes:
\eqn\simaction{
S=\int d^4x\ \sqrt{-g_E}\ \left(R_E - 2(\nabla\phi)^2 - e^{-2\phi}
F^2\right)}

When $F_{\mu\nu}=0$ this reduces to the standard
Einstein-scalar field action. The
``no hair" theorems \bekenstein\
show that the only black hole solutions of this
theory are Schwarzschild
with $\phi =0$ everywhere. Thus uncharged black holes in low energy string
theory are the same as general relativity.
Since this is simultaneously the simplest and most
physical black hole solution, it is extremely important to find
the corresponding exact conformal field theory. The exact solution should agree
with Schwarzschild until the curvature becomes of order the Planck scale. As
we have remarked, for
black holes with mass much larger than the Planck mass, this is well within
the horizon.
This shows that string theory has black hole solutions.

Since  the dilaton $\phi$
couples to $F^2$, charged black holes are {\it not} \RN\
with $\phi=0$.  One might worry that with the exponential coupling,
the exact solutions would be very complicated.  But, as we have seen, they are
are  easily found using the transformation
\trans.
(Recall that this transformation yields the
string metric.) Starting with a Schwarzschild solution with mass $m$
(and radial coordinate $\r$) one obtains
\foot{This solution was first found by
Gibbons \gibbons,
and further discussed in \gima. It was independently found a few years
later in a somewhat
simpler form by Garfinkle et.al. \ghs. All of these papers directly solved
the field equations. The solution generating technique described
here was discovered more recently.}
$${ ds^2 = -\(1-{2m \over \r}\)  \(1+{2m \ss\over \r}\)^{-2} dt^2 +
     \(1-{2m\over \r}\)^{-1} d\r^2 +\r^2 d\Omega }$$

$$ A_t = -{m \sinh 2\aa \over \sqrt 2 [\r+ 2m\ss]} $$

\eqn\strn{ e^{-2\phi} = 1+ {2m\over \r}\ss }
The causal structure of this spacetime is identical to Schwarzschild. There
is an event horizon at $\r=2m$ and a curvature singularity at $\r=0$. (The
vector potential $A_t$
is actually finite at $\r=0$, although the invariant $F_{\mu\nu}
F^{\mu\nu}$ diverges there.) Notice that unlike Schwarzschild, $g_{tt}$
vanishes
at the singularity as well as the horizon.
Of particular interest is the fact that there is no inner horizon.  I used to
think this is was a result of the instability of the inner horizon: When
the dilaton is included the inner horizon becomes singular.
But as we will see in the next sections, there are several examples of
solutions
with dilaton which have a nonsingular inner
horizon.

As $\r\a 0 $, $e^\phi\rightarrow 0$, so the string
coupling is becoming very {\it weak} near the singularity.
As we have discussed,
we have no right to trust this solution near the singularity,
but its difficult to resist speculating about what it might mean if the
exact classical solution had a similar behavior.  It would suggest that,
contrary to the usual picture of large quantum fluctuations and spacetime
foam near the singularity, quantum effects might actually be
suppressed. The singularity would behave classically\foot{This is not apparent
when the action is expressed in  terms of
the Einstein metric because Newton's constant has been suppressed. In
string theory, Newton's constant is not fundamental, but determined by
the dilaton and the string tension.}!

The physical mass $M$ of a solution to \fdeq{}\  is independent of whether
one expresses it in terms of the string metric or Einstein metric. Although
the formula for the mass in terms of the asymptotic form of the metric
and dilaton does depend on this choice. The easiest way to calculate the
mass of \strn\ is to rescale to the Einstein metric and compare with
Schwarzschild asymptotically. The physical charge $Q$ is of course related
to the asymptotic form of $A_t$. One finds that $M$ and $Q$ are related to
the original Schwarzschild mass $m$ and  transformation parameter $\aa$ by
\eqn\mqsol{ M = m \cosh^2 \aa, \quad Q = \sqrt 2 m \cosh\aa \sinh\aa  }
It follows that the charge to mass ratio depends only on $\aa$ and is given by
$Q^2/M^2 = 2 \tanh \aa $. Thus for fixed $M$, one can increase the charge by
increasing $\aa$ and decreasing $m$. This results in the area of the event
horizon becoming smaller.

\ifig\tnullsing{The extremal black hole with dilaton.}{\epsfbox{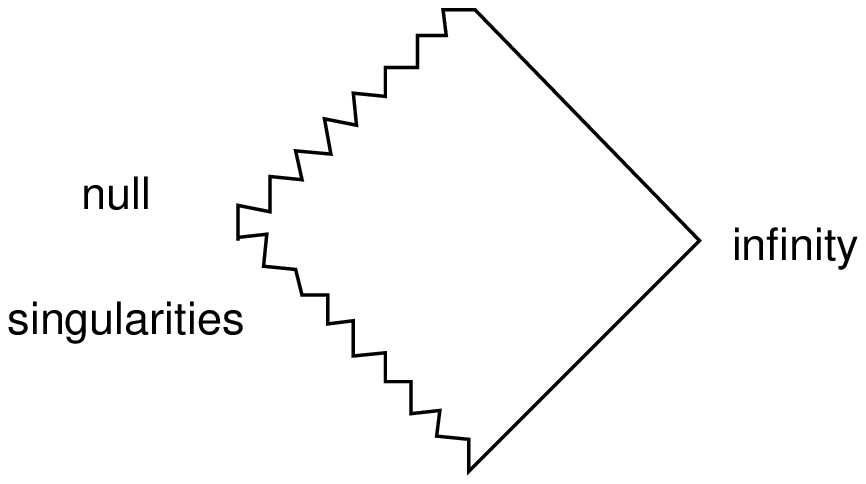}}

The largest possible charge for a given mass is $Q^2 = 2M^2$ and
is obtained by taking the limit $m \a 0 ,\aa \a \infty$ keeping $m\cosh^2 \aa$
fixed.
In this limit, the event horizon shrinks to zero size and becomes singular.
The metric takes the  extremely simple form:
\eqn\eextremal{
ds^2 = -\left(1+\frac{2M}{\r}\right)^{-2} dt^2 +
d\r^2 + \r^2 d\Omega}
The spatial part of the metric is now completely flat! Strictly speaking
this spacetime does not represent a black hole since it does not possess a
regular event horizon. Nevertheless, in analogy with the \RN\
solution,
we will call it the extremal charged
black hole. The singularity at $\r=0$ is not a typical naked singularity
like the one shown in \tnaksing.
It actually consists of two parts, each of which is
null. The Penrose diagram is shown in \tnullsing.
This can be understood as follows.
Near the singularity, radial null geodesics satisfy $\pm dt \propto  d\r/\r$
which implies that as $\r \a 0$,
the geodesics reach arbitrarily large values of $|t|$. This shows that
an outgoing null geodesic must cross every ingoing null geodesic.
Notice that the
condition for the extremal limit
has shifted from $Q^2=M^2$ (without the dilaton)
to $Q^2=2M^2$.  We will see a way to understand this in Sec. 4.4.

To facilitate comparison with the standard black holes of general relativity,
it is convenient to rescale to the Einstein metric. Performing this rescaling
and introducing a new radial coordinate $r = \r+2m \sinh^2 \aa$
yields a remarkably simple form of the solution:
$$
ds^2_E = - \(1-\frac{2M}{r}\) dt^2 + \(1-\frac{2M}{r}\)^{-1} dr^2
+ r\(r-\frac{Q^2}
{M} \)d\Omega $$
\eqn\einmet{ F_{rt}=\frac{Q}{r^2}\qquad e^{2\phi}
= 1-\frac{Q^2}{Mr}}
Note that the metric in the $r-t$
plane is identical to Schwarzschild!  The only difference is
that the area
of the spheres is smaller.
In fact, this area goes to zero when $r=Q^2/M$ and
this surface is singular. Since $g_{tt}$ remains finite at the singularity,
there
is no ``infinite stretching" analogous to what happens to an observer hitting
the singularity in Schwarzschild.
The Penrose diagram (for $Q^2 < 2M^2$)
is identical to
Schwarzschild.

As you increase $Q$, the singularity moves out in ``$r$''.  In the extremal
limit $Q^2 = 2M^2$, the singularity coincides with the horizon. Since the
causal structure in the $r-t$ plane is independent of $Q$, this shows that
the spacetime is described by \tnullsing\ even in the Einstein metric.
If you increase $Q$ farther, the
singularity moves outside the horizon and becomes timelike.

The fact that the horizon shrinks to zero size in the extremal limit has
an important consequence. It shows that there is no classical process
which will cause a
nearly extremal black hole  to become extremal \garfinkle.
This is a result of the area theorem \hawking\
which states that one cannot classically
decrease the area of a black hole. This theorem depends on an energy condition
which is indeed satisfied by \simaction.

So far we have discussed only electrically charged black holes.
The magnetically charged black hole can be obtained from the
electrically charged solution by an electromagnetic duality transformation.
{}From \simaction, the equation of
motion for $F_{\mu\nu}$ is
\eqn\maxwell{
\nabla_\mu\left(e^{-2\phi} F^{\mu\nu}\right) = 0}
This implies $\tilde F_{\mu\nu} \equiv e^{-2\phi}\half
{\epsilon_{\mu\nu}}^{\rho\sigma}F_{\rho\sigma}$ is curl free.
One can easily check that the
equations of motion are invariant under $F \rightarrow \tilde F$,
$\phi\rightarrow -\phi$, and $g_E\rightarrow g_E$. Starting with the
electrically charged solution and applying this transformation, yields the
magnetically
charged solutions.  Since the Einstein metric is unchanged, the Penrose
diagrams
are unchanged. However, since $\phi$ changes sign, the string coupling
becomes {\it strong} near the singularity for these black holes.

Let us now consider the magnetically charged black hole in terms of the
string metric.  The Einstein metric doesn't
change when we go from electric to magnetic charged black hole, but
since $\phi$ changes sign, the string metric {\it does} change. We get,
\eqn\mstringbh{
ds^2 =
-\frac{\left(1-\frac{2M}{r}\right)}{\left(1-\frac{Q^2}{Mr}\right)} dt^2
+ \frac{dr^2}{\left(1-\frac{2M}{r}\right)\left(1-\frac{Q^2}{Mr}\right)}
+ r^2 d\Omega}
As one approaches the singularity $r=Q^2/M$, the area of the two spheres does
not go to zero. Thus in this sense the singularity is larger than the previous
case.
Now the extremal limit is:
\eqn\mextremal{
ds^2 = -dt^2 + \left(1-\frac{2M}{r}\right)^{-2} dr^2 + r^2 d\Omega}
{\it This metric (with $r>2M$) is geodesically complete and
has no curvature singularities!}
A $t=$const
surface is identical to a  $t=$const surface in  extreme \RN\
and resembles an infinite throat.
There is an infinite
proper distance from $r=2M$ to any larger  value of $r$.
 But unlike \RN, $r=2M$ is now also infinitely far away
in null and timelike directions.  The horizon has moved off to infinity
taking the singularity with it. The dilaton is $e^{-2\phi} =
1-\frac{Q^2}{Mr} = 1-\frac{2M}{r}$ for the extremal case.
In terms of proper distance along the infinite throat,
$\rho \equiv 2M ln (r-2M),$ we have $\phi = -\rho/4M$. So the
dilaton is {\it
linear} as $\rho\rightarrow - \infty$ and rolls to strong coupling.

Linearized perturbations around these black holes have been studied \howi.
One finds that they are classically stable, just like the more
familiar Einstein-Maxwell
black holes. It has also been shown that the extremal black holes are
supersymmetric when embedded in an $N=4$ supergravity theory \klopp.

The Hawking temperature of these black holes is easily determined
as follows.
Rescaling the metric by a function which is smooth and nonzero at the horizon,
and goes to
one at infinity does not affect $T_H$. So one can calculate the temperature
using the Einstein metric.  But the temperature only depends on the metric
in the  $r-t$ plane, and for this part of the metric $g_E$ is identical to
Schwarzschild. Thus the Hawking temperature is same as Schwarzschild $T_H =
1/8\pi M$ {\it independent of $Q$}! Thus unlike \RN,
$T_H$ does not vanish
in the extremal limit.  This  leads one to worry that a black hole
may keep radiating past the extremal limit
and become a naked singularity.

There are at least two ways to avoid this conclusion. The first is if there are
large potential barriers outside the horizon which cause most of the radiation
to be reflected back. If these barriers exist, and
grow as the extremal limit is approached,
the amount of energy radiated to infinity could go to zero and not result in a
naked singularity. It turns out that there are potential barriers but they
are not large enough, by themselves, to prevent the formation
of naked singularities \howi\gist.
The second possibility is to include back reaction. Since the horizon moves
from a finite to an infinite distance in the string metric,
it is clear that the classical geometry
is changing significantly as one approaches the extremal
limit. This must be properly taken into account. Also, it has been shown
that the thermal approximation breaks down  near the extremal limit \psstw.
A complete calculation has not yet been done.
A model two dimensional problem including backreaction is currently
under investigation.
(For a recent review, see \hast.)

Although the temperature does not go to zero in the extremal limit, the
extremal black hole itself has zero temperature. This is because it has no
event horizon and is globally static. The analytic continuation to Euclidean
space does not require periodically identifying imaginary time. Thus the
Hawking temperature is discontinuous in this limit, which is further evidence
that the thermal approximation is breaking down.

\newsec{MORE GENERAL BLACK HOLE SOLUTIONS}

In this section we will consider several generalizations of the basic black
hole solution discussed above.  These will include adding rotation,
electric and magnetic charges, higher dimensions, etc. To keep the
discussion manageable, we will consider these generalizations independently,
starting each time with the black hole solution in Sec. 3. If desired, one can
construct even more general solutions by combining several
of these features simultaneously.

\subsec{Nonzero Dilaton at Infinity}

The solutions discussed so far all have vanishing dilaton at infinity.
Since the value of the dilaton at infinity determines the string coupling
constant at large distances from the black hole, one might wish to keep
this a free parameter. Fortunately, it is easy to extend these solutions
to allow $\phi$ to approach an arbitrary constant $\phi_0$.
The action \simaction\
is clearly invariant under $g_E \rightarrow
g_E, \phi\rightarrow \phi + \phi_0, F\rightarrow e^{\phi_0} F$.
Even though the Einstein metric is invariant under this transformation, when
expressed in terms of the physical charge, it will depend on $\phi_0$. This
is simply because the charge is rescaled by $e^{\phi_0}$.
(One can obtain the same solution by keeping the Maxwell field fixed and
rescaling the Einstein metric.)
For example, applying this to the magnetically charged black hole,
the solution becomes
$$ds^2_E = - \(1-\frac{2M}{r}\) dt^2 + \(1-\frac{2M}{r}\)^{-1} dr^2
+ r\(r-\frac{Q^2 e^{-2\phi_0}}
{M} \)d\Omega$$
\eqn\phinz{ F_{\theta\varphi}=Q\sin \theta \qquad e^{-2\phi}
= e^{-2\phi_0}\(1-\frac{Q^2 e^{-2\phi_0}}{Mr}\)}
The extremal limit is now $Q^2 = 2M^2 e^{2\phi_0}$. So for large $\phi_0$,
there exist black holes with $Q$ much larger than $M$.

\subsec{Both Electric and Magnetic Charges}

We have seen that in terms of the string metric, the
electric and magnetic black holes have very
different properties in the extremal limit. One has an infinite throat
and no curvature singularity,
while the other has a singularity but flat
spatial slices. What happens when both
charges are present?  One might think that since there is no reason to
prefer one type of charge over another, the extremal limit should
somehow combine the features of both. But that is not what happens. We will see
that the
magnetic charge dominates:  If $Q_M\not= 0$, the extremal limit
resembles the magnetic solution regardless of the value of $Q_E$!

In the Einstein-Maxwell theory, there is a continuous electromagnetic
duality rotation $ F\a \cos \theta F + \sin \theta\ {}^*F $
(where $*$ denotes the dual)
which
interpolates between electric and magnetic charges. Since the stress tensor
is left invariant, this will map solutions to solutions keeping the (Einstein)
metric invariant. In string theory, life is complicated by both the dilaton
and the antisymmetric tensor $H_{\mu\nu\rho}$.
(A solution containing both electric and magnetic charge must also contain
a nonzero $H_{\mu\nu\rho}$ because of the
fact that  $dH=-F\wedge F$.) Nevertheless,
there is a generalization of this electromagnetic
duality rotation which can be used to generate solutions with both
electric and magnetic charges\foot{A similar transformation had been noticed
earlier in the context of supergravity theories \csf.} \stw\senem.
As before, the Einstein metric is left invariant but
the charge $Q^2$ is now  interpreted as $Q^2\equiv Q_E^2 +Q^2_M$.
The dilaton is:
\eqn\emdil{
e^{2\phi} = \frac{1}{Q^2} \left[Q^2_E e^{2\tilde\phi} + Q^2_M
e^{-2\tilde\phi}\right]}
where
\eqn\newphi{e^{2\tilde\phi} = 1-\frac{Q^2}{Mr}}
Notice that $e^{2\phi}$ is simply the sum of its electric and magnetic
values.  This linearity is quite surprising since the field equations
are highly nonlinear.
The
fact that the magnetic charge dominates now follows immediately.
Near $r=Q^2/M, \ \ e^{2\tilde\phi}\rightarrow 0$.
Thus the contribution from $Q_E$ becomes negligible compared to that
from $Q_M$.  In particular, the string coupling will become strong near
the singularity whenever $Q_M\not= 0$, and the rescaled string metric
will be non-singular.

To complete the solution we must specify $F_{\mu\nu}$ and $H_{\mu\nu\rho}$.
The Maxwell field is simply the sum of the fields for
electric and magnetic charges.
In four dimensions, the antisymmetric tensor field
can be replaced by a scalar via
$H=- e^{4\phi}(*d\chi)$.
For the black hole solution,  the scalar is given by
\eqn\chisol{
\chi=Q_E Q_M \frac{e^{2\tilde\phi}-e^{-2\tilde\phi}}{Q^2_E e^{2\tilde\phi}
+ Q^2_M e^{-2\tilde\phi}}}
This vanishes when either $Q_E$ or $Q_M$ is zero as it should.
Since the Einstein metric is unchanged, the Hawking temperature is still
$T_H=1/8\pi M$ for these black holes.

The fact that the magnetic charge dominates is a consequence of the
three form $H$. If one ignores the $H$ field, black hole
solutions to \simaction\ which have both electric and magnetic charge, are more
symmetric in $Q_E$ and $Q_M$. The exact solution is known \gima\klopp\
and is
most conveniently
expressed in terms of the following parameters:
\eqn\xidef{  r_0 \equiv {Q_M^2 - Q_E^2 \over 2M}
\qquad  r_\pm \equiv M \pm (M^2 + r_0^2 - Q_E^2 - Q_M^2)^\half}
The Einstein metric is
then
\eqn\symsol{ ds^2_E = - {(r-r_+)(r-r_-)\over r^2 - r_0^2} dt^2
	+ {r^2 - r_0^2 \over (r-r_+)(r-r_-)} dr^2 + (r^2 - r_0^2) d\Omega}
and the dilaton is
\eqn\embkdil{ e^{2\phi} = {r+r_0 \over r-r_0} }
This solution is invariant under interchanging the charges $Q_M$ and $Q_E$
and changing
the sign of $\phi$.
Near the singularity,
the dilaton goes to strong or weak coupling depending on which of the two
charges is larger.
As expected, if $Q_E$ or $Q_M$ vanish, we recover our previous dilaton
black hole solutions \einmet\ (with the radial coordinate shifted by $r_0$).
If $Q_E = Q_M$, then the dilaton vanishes
and the metric reduces to the familiar \RN\ solution. This is also what
one should expect since the source of the dilaton is proportional to
$F^2$ which vanishes when $Q_E = Q_M$. Thus \symsol\ provides
an interesting interpolation between these two classes of solutions.

When $Q_E$ and $Q_M$ are both nonzero, the global structure of \symsol\
is similar to \RN\ with
an event horizon at $r_+$, an inner horizon at $r_-$,
and a curvature singularity
at $r=|r_0|$. This is our first example of a solution with an inner horizon
and a nontrivial dilaton. In the extremal limit,
the two horizons coincide and the
temperature vanishes. If one of the charges
vanish, the extremal black hole now has a temperature which depends on the rate
at which the charge goes to zero as  the extremal hole is approached. Any
temperature between $0$ and $1/8\pi M$ can be obtained.

Although we have ignored the antisymmetric tensor $H$,
the solution \symsol\
can still be viewed as a solution to low energy string theory
if we interpret the charges as being associated with two different
Maxwell fields $F^i_{\mu\nu}$.
Since string theory starts with a large gauge group, it is certainly possible
to have two
unbroken $U(1)$ subgroups. In this case, the source of $H$ would be $F^i \wedge
F_i$ which vanishes. (Solutions where both $U(1)$ fields have electric
and magnetic charges, and $H$ is nonzero have recently been constructed
\ortin.)

\subsec{Rotation}

So far we have considered only non-rotating
black holes.  When the charge is zero, the rotating black hole in string
theory is the same as general relativity and is called the Kerr
solution:
\eqn\kerr{\eqalign{
ds^2_E = & - \left(1 - {2mr \over \Sigma} \right) \, dt^2
       + { \Sigma \over \Delta} \, dr^2 + \Sigma \, d\theta^2
   - { 4mr a \sin^2 \theta  \over \Sigma} \,dt \, d\varphi \cr
    & + \left[ {(r^2 + a^2)^2 - \Delta a^2 \sin^2 \theta \over \Sigma} \right]
					    \sin^2 \theta \, d\varphi^2 }}
where
\eqn\sigmadef{ \Sigma = r^2 + a^2 \cos^2 \theta \>, }
\eqn\deltadef{ \Delta = r^2 + a^2  - 2 m r \>, }
and $a$ is the angular momentum $J$ divided by the mass.
This solution is similar to \RN\ in terms of its causal structure.
When $|a|< m$, there are two horizons at the zeros of $\Delta$.
When $|a|=m$, these horizons coincide,
 and
when $|a|>m$, they disappear and the spacetime contains a
naked singularity. Also, like \RN, in the extremal limit
$(|a| = m)$ the Hawking temperature $T_H$ vanishes and the event horizon
remains nonsingular with non-zero area $A$.
This behavior is quite different from the extremal limit
of the charged non-rotating black hole.   We saw in Sec. 3
that as one approaches the extremal limit  ($Q^2 = 2M^2$) the event horizon
in the Einstein metric becomes singular:
$A\a 0$ and the dilaton diverges there.
In addition, $T_H\rightarrow 1/8\pi M$. Now consider a black hole which has
both charge and rotation.  We again have a situation like Sec. 4.2
where two
special cases have different extremal limits.  What is the behavior of
the general extremal black hole  with both charge and rotation in string
theory?
It turns out that angular momentum dominates over charge. If $J\ne 0$,
then the extremal limit resembles the Kerr solution, independent of the
value of $Q$.

The solution for a rotating charged black hole in string theory was found
by Sen~\senrot, by applying a  generalization of the transformation \trans\
to the Kerr
solution.
Since a rotating charged
black hole has a magnetic dipole moment, $F\wedge F \ne 0 $. So again one
must include $H$. In terms of the Einstein metric, the solution is
\eqn\rotbh{\eqalign{ ds^2_E =
& -\left(1-{2mr\cosh^2\aa \over \Upsilon}\right)dt^2
+ {\Upsilon \over \Delta} dr^2 + \Upsilon
    d\theta^2  - {4mra\cosh^2\aa \sin^2 \theta \over \Upsilon}
   dt d\varphi \cr &+ \[{(r^2 + a^2 +2mr\sinh^2\aa)^2 -\Delta a^2\sin^2\theta
     \over \Upsilon} \]\sin^2\theta  d\varphi^2}}
where
\eqn\updef{ \Upsilon = r^2 + a^2\cos^2 \theta + 2mr\sinh^2\aa  }
and $\Delta$ is defined as before \deltadef.
This is essentially the same as the Kerr metric with $\Sigma$ replaced
by $\Upsilon$.
The Maxwell field, dilaton, and antisymmetric tensor are
$$ A= -{mr \sinh 2\alpha \over \sqrt 2\Upsilon}
(dt - a \sin^2\theta d\varphi) $$
$$  e^{-2\phi} = {\Upsilon \over \Sigma}$$
\eqn\abdef{  B_{t\varphi} = {2mra\sinh^2\aa \sin^2 \theta \over \Upsilon} }
One can easily verify that this solution has the correct limits when $\aa \a 0$
or $a \a 0$.
The mass $M$ and charge $Q$ are related to $m$ and $\alpha$ in exactly the
same way as the nonrotating solution \mqsol, and  the angular momentum
is given by $J=Ma$.
This solution has two horizons (at the zeros of $\Delta$)
when $m > a$ which corresponds to $2M^2 > Q^2 +2|J|$.
The presence of rotation does not have much affect on the behavior of the
dilaton.
The string coupling $e^\phi$ still goes to zero at the singularity $\Sigma = 0$
as  expected for an electrically charged black hole.

The extremal limit is $2M^2 = Q^2 +2|J|$. (Since this corresponds to $m= a$,
the transformation generating this solution  preserves the extremality of
the black hole when it adds charge.) In this
limit, one can show that
the area of the event horizon is simply related to the angular momentum
\eqn\area{ A=8\pi|J|}
and is independent of $Q$.
This clearly shows how the zero area of the nonrotating black hole is modified
by rotation. When $J$ is nonzero, the horizon is perfectly regular in the
extremal limit. In
particular, the dilaton remains finite there. If one does a duality
rotation to obtain a rotating magnetically charged black hole, the situation
is similar. This shows that the string metric will be qualitatively the
same as the Einstein metric. It will {\it not} have
an infinite throat. Thus the ``generic" black hole resembles Kerr.
One can also show that the Hawking temperature goes to zero in the extremal
limit whenever $J \ne 0$. However, since Hawking radiation carries away
angular momentum, it is possible that an evaporating black hole will approach
a nonrotating extremal limit.

A final comment about rotating black holes concerns the
gyro-magnetic ratio.
A rotating charged black hole has a magnetic dipole moment $\mu$ so
one can compute a $g$-factor from the ratio  $\mu/J$. For the Einstein-Maxwell
theory, one has the remarkable result that black holes have $g=2$
(the value for electrons), rather than $g=1$ which one might have expected
since
this is the value for classical matter.  It turns out that string
black holes also have $g=2$. One might be tempted to
extrapolate from this that all black
holes have $g=2$, but this is incorrect. For instance, suppose one simply
leaves out the $H$ field and considers rotating charged
black hole solutions to \simaction.  Then
one finds that $g$ depends on the charge to mass
ratio of the hole and varies from $g=2$ for small charge to $g=3/2$ in the
extremal limit\foot{Black holes in Kaluza-Klein theory behave
similarly
\giwi.}\hohorot. The value $g=2$ is recovered only when $H$ is included
in exactly the manor predicted by string theory. The significance of this
is not yet understood.

\subsec{Multi-Black Holes}

For extremal \RN\ black holes,
the gravitational attraction exactly balances the electromagnetic repulsion,
and there exist static multi-black hole solutions.  The same is true for
the solutions described in Sec. 3.
In fact, one can understand the fact that the extremal
limit $Q^2=2M^2$ has a larger charge/mass ratio than general relativity as a
result of the fact that the dilaton contributes an extra attractive
force.  So, for a given $M$ one needs a larger $Q$ to balance it. More
explicitly, for a static
asymptotically flat solution, one can define a dilaton charge
\eqn\dilchg{  D = {1\over 4\pi} \int d^2 S^\mu \d_\mu \phi }
where the integral is over the two-sphere at infinity.
In agreement with ``no hair" theorems, this dilaton charge is not an
independent
free parameter but is uniquely determined by the mass and charge. For the
charged black hole solution \einmet, one finds
$D= Q^2/2M$.
For two widely separated black holes with mass and charge $M_i, Q_i$, the
total force is thus
\eqn\force{\eqalign{
{\cal F} = &\[Q_1 Q_2 - M_1M_2 -{Q^2_1 Q^2_2 \over 4M_1M_2}\]{1\over r^2} \cr
   = &-{M_1M_2 \over 4r^2} [{Q_1Q_2 \over M_1M_2} -2]^2 }}
So the force vanishes when $Q_1Q_2 = 2M_1M_2.$
For black holes, $Q_i \le \sqrt 2 M_i$
and the force vanishes in the extremal limit.

If
$\vec x$ are Cartesian coordinates on ${\bf R}^3$, then the solution
describing a collection of extremal electrically charged
black holes of mass $M_i$ located at
$\vec x_i$ is
\eqn\emultbh{\eqalign{
ds^2 = &-e^{4\phi} dt^2 + d\vec x \cdot d \vec x \cr
   e^{-2\phi} =& 1 + \Sigma_i
\frac{2M_i}{|\vec x - \vec x_i|}}}
This clearly reduces to the previous result \eextremal\ for one black hole.
Space is
again completely flat and there are singularities at the location of each
black hole.
For magnetically charged extremal black holes the solution is
\eqn\mmultbh{\eqalign{  ds^2 =& -dt^2 + e^{4\phi}d\vec x \cdot d \vec x \cr
   e^{2\phi} =& 1 + \Sigma_i
\frac{2M_i}{|\vec x - \vec x_i|}}}
Foe a single black hole, this is simply our previous solution \mextremal\
reexpressed in isotropic coordinates.
A spatial surface now  looks like $\R^3$ with a finite number of  throats
branching off.
The Maxwell field in each case is just the sum of the Maxwell fields for
single black holes.

\subsec{Massive Dilatons}

In all the black hole solutions we have discussed so far, the dilaton was
assumed to be strictly massless. While this is the prediction of classical
low energy string theory, it is in conflict with experiment. In many respects
the dilaton acts like a Brans-Dicke scalar, with a coupling that violates
observational limits. Fortunately, there are strong theoretical arguments
that the dilaton should have a mass. The dilaton must be massless as long as
supersymmetry is unbroken, but when supersymmetry is broken at low energy
it is likely to acquire a mass. We are not yet able to do the nonperturbative
quantum calculations required to calculate the low energy dilaton potential.
We will consider the simplest choice\foot{For a discussion of
black holes with more general dilaton potentials, see~\grha.}\ $m^2 \phi^2$.

The black hole solutions with a massive dilaton
do not appear to be expressible in closed form.
However by combining approximate solutions with numerical results one can
obtain a fairly complete picture of their properties \hohomass. They differ
from
the massless dilaton solutions in several respects. At large distances,
the dilaton now falls off like $1/r^4$. This causes the dilaton contribution
to the stress tensor to be negligible compared to the  charge terms. Thus,
at large distances, the solution always approaches \RN. However the presence
of the dilaton near the horizon still allows black holes to have $Q^2>M^2$.
Since the dilaton force is negligible at large distances, {\it nearly
extremal and extremal black holes repel each other}. This may be the first
example of gravitationally bound
repulsive objects. One can show that for a large
black hole, the extremal limit corresponds to $Q^2 = M^2 + 1/5m^2$.
The energy of $n$ widely separated black holes with the same total charge is
then
\eqn\bifur{ M_n = n\[\({Q\over n}\)^2 - {1\over 5m^2} \]^{1\over 2}
 = \[Q^2 - {n^2\over 5m^2} \]^{1\over 2}  \>. }
Clearly, $M_n$ is a decreasing function of $n$. In other words, when the
dilaton is massive, it is energetically favorable for a large extremal
black hole to split into several smaller black holes. This process cannot
occur classically, but presumably can occur quantum mechanically.

The causal structure of extremal black holes in this theory depends on their
charge. If $|Qm| > e/2$, then
the extremal limit is similar to \RN. There are two horizons which
come together. This corresponds to an extremal black hole which is larger
than the Compton wavelength of the dilaton. If $|Qm|<e/2$ then there is only
one horizon. The extremal limit is then similar to the massless dilaton
black holes.  In particular, the string metric describing the extremal limit
of a magnetically charged black hole will have an infinite throat.
Physically, a large black hole with small charge, will start off close to the
\RN\ solution, but as it evaporates, it will begin to resemble the massless
dilaton solution. If the mass of the dilaton is about 1TeV, the transition
will occur when the black hole has a mass about $10^{11}$ gms. This is
sufficiently large that other string corrections should still be negligible.

Another unusual property of black holes with a massive dilaton is the
following. It is well known that the maximally extended
Schwarzschild solution has a wormhole
in the sense that a spacelike surface connecting the two asymptotically
flat regions reaches a minimum size inside (or on) the black hole.
However this wormhole cannot be transversed, since it quickly collapses to
zero size. One can show \hohomass\
that the string metric describing certain black holes
coupled to a massive dilaton have a wormhole outside the horizon. This wormhole
is static and can be transversed. This only occurs when the size of the
hole is of order the Compton wavelength of the dilaton. (In this case there
is a slight possibility that the black hole solution will have {\it three}
horizons \grha.)

\subsec{Lower Dimensions}

There is no two dimensional analog of the Schwarzschild solution in
general relativity for the simple reason that Einstein's equation becomes
trivial. However the low energy
string action \action\ is nontrivial even in two dimensions.
(Although in this case, one can  no longer rescale to the
standard Einstein action.) Of course, the three form $H$ must vanish
in two dimensions.
Let us further assume that $F=0$ to begin.
Naive counting
indicates that gravity in two dimensions has $-1$
degrees of freedom. This suggests that gravity plus the dilaton should have
zero degrees of freedom. While this indicates
that there are no propagating modes,
there can still be nontrivial solutions. In fact,
there are black hole solutions to \fdeq{}\
provided one includes the constant $\Lambda$ and allows the dilaton to
grow linearly at infinity. The solution takes the form \msw\witten
\eqn\lowbh{\eqalign{ ds^2 =& -\left(1 - {M \over r} \right) d t^2 +
	    { k dr^2\over 4 r (r - M)}  \cr
 \phi =& -{1\over 2}\ln r - {1\over 4}\ln k  }}
where  $M$ is the mass and $k$ is related to the constant $\Lambda$.
One can show that these are the only  classical solutions
in two dimensions. To obtain charged black holes one can simply apply the
transformation \trans\ \ils.

What is the point of studying black holes in two dimensions with unusual
boundary conditions, when more physical solutions are known in four dimensions?
The answer is that by going to two dimensions, one can progress much farther
than simply solving the low energy string equations of motion. For example,
Witten has found the corresponding exact conformal field theory~\witten. This
is
obtained by starting with  a Wess-Zumino-Witten (WZW)
model based on the noncompact
group
$SL(2,\R)$ and gauging a one dimensional subgroup.
One intriguing feature of the exact conformal field theory is that it includes
a region of spacetime
``beyond" the singularity corresponding to $r<0$ in \lowbh.
However it appears unlikely that strings could propagate through the
singularity. If so, there are potential causality problems since the
light cones tip over
on the other side. (Actually, the exact conformal field
theory contains two copies of the entire maximally extended black hole
spacetime
which are joined at the singularity. It is not clear whether this has
any physical significance.) In the supersymmetric case, the only higher order
corrections to the solution \lowbh\ is an overall rescaling of the metric
\jjp\basf. In the purely bosonic case, there are other corrections.
The  corresponding exact metric has been
found~\dvv\basf\  and is given by
\eqn\exactbh{ds^2 =  -\beta^{-1}(r) \(1-{M\over r}\) dt^2
    +{(k-2) dr^2 \over 4r(r-M)}}
where
\eqn\betadef{ \beta(r) = 1- {2\over k} \(1-{M\over r}\)}
This agrees with the above metric for large $k$ which is equivalent to
small curvature (since, by rescaling $t$, one can view $k$ as multiplying the
entire metric).
The metric is now regular at the former singularity $r=0$.
But it still has a curvature singularity at a negative value of $r$
where $\beta(r) = 0$. The exact dilaton has also been calculated and is
\eqn\exactdil{\phi = -{1\over 2}\ln r\sqrt{\beta(r)} +\phi_0 }
where $\phi_0$ is a constant.
Notice that this still diverges at the original location of the singularity
$r=0$.

Another advantage of two dimensions is that one can study Hawking evaporation.
One would like to do this in the context of  the full quantum
string theory,  perhaps taking
advantage of the recent progress in nonperturbative solutions of string theory
in two dimensions. However a more modest goal is to add matter to the
low energy string
action \action\ and study Hawking evaporation in this theory. This has been the
subject of extensive work over the past year and is reviewed in
the lectures by Harvey and Strominger in this volume
 \hast.

What about three dimensional black holes? Given the two dimensional black hole,
one can clearly take its product with $S^1$ to obtain a three dimensional
solution with an event horizon.  Unfortunately, this is the best one can do.
There do not exist any other static, axisymmetric
three dimensional solutions of \fdeq{}\ with a regular horizon. This can be
seen
as follows\footnote*{This argument was developed in collaboration
with J. Horne.}. For simplicity, we set $F=H=0$. (If a charged black hole
exists in three dimensions, then an uncharged one should exist as well.)
The equation for the metric \fdeq{a} is
\eqn\ricci{    R_{\mu\nu} = -2 \d_\mu\d_\nu \phi }
Let us assume a metric of the form
\eqn\genmet{   ds^2 = -\lambda dt^2 + fdr^2 + gd\theta^2}
By combining the $tt$ and $\theta\theta$ components of \ricci\ one obtains
\eqn\equa{{\lambda'' \over \lambda'} - {\lambda' \over \lambda} = {g'' \over
g'} -
	 {g' \over g} }
This equation is immediately integrated to yield:
\eqn\soln{    g = c_1 \lambda^{c_2}}
where $c_1$ and $c_2$ are constants. For a regular horizon, one needs
$\lambda = 0$ with $g$ remaining finite and nonzero. This is possible only if
$c_2 = 0 $ which yields the simple product of a two dimensional solution with
$S^1$. Although the curvature of this solution vanishes at infinity, it
is not asymptotically flat in the usual sense of approaching the flat
metric on $\R^2$ minus a ball. We are forced to conclude that there are
no asymptotically flat three dimensional black holes in string theory.

Notice that this argument is independent of any boundary conditions
on the dilaton
at infinity. Since there are black hole solutions with the dilaton growing
linearly at infinity in two dimensions, and going to a constant in four
dimensions, one might have thought that there would be solutions with
$\phi$ growing, say, logarithmically in three dimensions. This argument
shows that such solutions do not exist.

It has recently been shown that there {\it are} three dimensional black holes
in general relativity with negative cosmological constant~\btz.
It is not yet clear whether they have any significance for string theory.

\subsec{Higher Dimensions}

There is a straightforward
generalization of the electrically charged black hole to higher
dimensions \gima.
One simply starts with the $D$ dimensional Schwarzschild solution
and
applies the transformation \trans.
Since the $D$ dimensional Schwarzschild solution
is related to the four dimensional solution by essentially
replacing $r$ by $r^n$ where $n=D-3$, the same is true for the stringy version
\eqna\hdbh
$$ \eqalignno{ds^2 = &-\(1-{cm\over r^n}\)
   \(1+{cm\ss \over r^n }\)^{-2} dt^2 \cr
    & +\(1-{cm\over r^n}\)^{-1}dr^2 +r^2 d\Omega_{n+1}   &\hdbh a \cr
 A_t =&- {cm \sinh 2\aa \over 2 \sqrt 2 [r^n+ cm\ss]} &\hdbh b \cr
 e^{-2\phi} =& 1+ {cm\over r^n}\ss   & \hdbh c }$$
 where $c$ is a dimension dependent constant.
The  mass and charge  are given by
\eqn\mqhd{M=m\(1+{2n\over n+1}\ss\)  \qquad Q = cmn \c\s /\sqrt 2}
To obtain the Einstein metric one multiplies \hdbh{a}\  by $e^{-4\phi/D-2}$.
These solutions have an
event horizon at $r^n=cm$ and a singularity at $r=0$.
In the extremal limit the horizon shrinks to zero size and the spatial
metric becomes flat.  Note that the  string coupling vanishes,
$g=e^\phi\rightarrow 0$ near the singularity in all dimensions.

Although
the higher dimensional black holes resemble the four-dimension black holes
in almost all respects, there are two important differences. In the extremal
limit, the singularity is no longer null, but is now timelike (like
the negative mass Schwarzschild solution). This can be seen from the fact
that radial null rays satisfy $\pm dt \propto dr/r^n $. For $n>1$, there is
only
a finite change in $t$ as $r \a 0 $. The second difference is with the
Hawking temperature. One can show from \gentemp\
that  $T_H\rightarrow 0$ as you approach the extremal
black hole for all $D>4$.  It is only for $D=4$ that the temperature
approaches a non-zero limit.

Unlike the electrically charged case,
there is {\it no} higher dimensional generalization
of the magnetically charged black hole for the simple reason that there is no
magnetic charge in higher dimensions. $Q_M$ is defined by
integrating $F$ over the sphere at infinity and only in four spacetime
dimensions is the sphere at infinity two dimensional.
(The electric charge does not have a similar problem
 since $Q_E \propto \int_{S_\infty} {}^* F$ and $^*F$ is a $D-2$ form
which can be integrated over the $D-2$ sphere at infinity for all $D$.)

Even though there are no black holes with magnetic charge in higher
dimensions, for $D=5$ there is an analogous solution using the three form
$H$. One can define a charge associated with $H$ in exactly the same way
one defines magnetic charge in four dimensions:
$\hat \Q = \int_{S_\infty} H / V_3$.
The five-dimension black
hole with $\hat \Q \not= 0$ is \host
$$
ds^2= - \frac{\left[1-(r_+/r)^2\right]}{\left[1-(r_-/r)^2\right]}\ dt^2
+ \frac{dr^2}{\left[1-(r_+/r)^2\right]\left[1-(r_-/r)^2\right]} + r^2
d\Omega_3 $$
\eqn\fivebh{  \ e^{-2\phi} = 1-(r_-/r)^2\qquad H=\hat \Q\epsilon_3 }
where
$\epsilon_3 $ is the volume form on a unit three sphere.
The constants $r_+, r_-$ are related to $M, \hat \Q$ by
$$
M= r^2_+ - \frac{1}{3} r^2_-, \qquad \hat \Q=2r_+r_-
$$
This is very similar to the four-dimensional magnetically charged
solution\foot{
I know of no way to obtain it by a transformation of Schwarzschild.
In five dimensions, the dual of $H$ is a two form, but it has a different
coupling to the dilaton than the Maxwell field $F$. Thus one cannot obtain
\fivebh\ as we did in four dimensions, by dualizing the electrically charged
black hole. The above solution was found by explicitly solving the
field equations.} \mstringbh.
The event horizon is at $r=r_+$ and the singularity is at $r=r_-$.
The extremal limit
($r_+=r_-$) is again completely nonsingular:
\eqn\fiveext{
ds^2 = - dt^2 + \left[1-\left(\frac{r_+}{r}\right)^2\right]^{-2} dr^2
+ r^2 d\Omega_3}
As in the previous case \mstringbh,
the dilaton is again linear in proper distance along the throat
and rolls to strong coupling.
One can show \chs\
that this extremal limit is in fact an exact solution of the
type II superstring theory. (To get a solution to the
heterotic string one must add appropriate gauge fields.)

There is, in fact, a second way to take the extremal limit, in which
one stays a finite distance from the horizon as the limit is taken \gistexact.
The
asymptotic region is now lost, and the solution becomes just a product
of the two dimensional black hole with $S^3$. This is also an exact conformal
field theory since the $H$ field has only components on the  $S^3$ and
is simply the SU(2) Wess-Zumino-Witten model.

This black hole has a magnetic type of $H$ charge. What about solutions
with electric $H$ charge i.e. $\Q \equiv \int_{S_\infty} {}^*H/V_{D-3}\not= 0$.
Since $^*H$ is a
$D-3$ form, it is
clear that this charge is carried by a one-dimensional extended
object i.e. a {\it string}.  A few years ago Dabholkar et. al.~\dghr\ found the
solution describing a straight fundamental string i.e. they added a
source term to the field equations \fdeq{}\
which was a $\delta$ function on the string worldsheet.  These
solutions had $\Q\ne 0 $ but
did not have an event horizon. We will see in Sec. 5
that there are solutions to \fdeq{}\ with $\Q\ne 0$ which describe
one-dimensional extended objects surrounded
by event horizons i.e. black strings.

\newsec{BLACK STRINGS}

 Black strings can be constructed in various spacetime
dimensions. As we will discuss in Sec. 5.3, for $D=3$ the exact conformal
field theory is known. However in
higher dimensions, we only know the solutions to the low energy field equations
\fdeq{}.

\subsec{Dimensions $D>4$}

Uncharged black strings in $D$ dimensions are simply the product
of a $D-1$ dimensional Schwarzschild solution and ${\bf R}$.
\eqn\unbksth{\eqalign{
   ds^2 = -\(1-{cm\over r^n}\)d\hat t^2 + &\(1-{cm\over r^n}\)^{-1}dr^2
	+r^2 d\Omega_{n+1}
	 + d \hat x^2
		      \cr
		 \phi = 0, &\qquad B =0 } }
where $n=D-4$ and $c$ is, as before, a dimension dependent constant.
Black strings with electric charge can similarly be obtained
by simply taking the product of the $D-1$ dimensional charged black hole and
${\bf R}$. It is more interesting to consider black strings with
$H$ charge. These are not simple products.
However, they can be found quite easily using spacetime
duality as described in Sec. 2.3. First we apply a Lorentz boost
$ \hat t = t\c + x\s, \hat x = x\c + t\s$, and then
a duality transformation \sigmadual\ on $x$ to get \host\hhs
\eqna\bksth
$$\eqalignno{
 ds^2 =& -{(1-cm/r^n)\over (1+cm\ss/r^n)}dt^2 +  {dx^2\over (1+cm\ss/r^n)} \cr
     &+{dr^2\over (1-cm/r^n)} + r^2 d\Omega_{n+1} & \bksth a \cr
	e^{-2 \phi} =& 1+{cm \ss\over r^n} &\bksth b \cr
	     B_{xt} =& {c m \c\s  \over r^n + cm\ss} \>
	     &\bksth c  } $$
Notice the similarity with the charged black hole solution \hdbh{}. The
dilatons
are identical, $B_{tx}$ is a multiple of $A_t$, and the only difference
between
the metrics is that the factor $(1+cm\ss/r^n)$ does not appear squared
in $g_{tt}$ but is split evenly between $g_{tt}$ and $g_{xx}$. As before,
the event horizon is at $r^n = cm$ and the curvature singularity is at $r=0$.
The metric is spherically symmetric, static, and translationally invariant
in $x$.
The causal structure of this dual
solution is exactly
the same as for Schwarzschild, as one might have expected since (when
$x$ is periodic) they represent the same conformal field theory.
However, like the charged black hole in string theory,
it is very different from \RN. In
particular there is no inner horizon and the singularity is not timelike.

The parameters $m$ and $\alpha$ are related to the physical mass and charge by
\eqn\bkstmq{M= m(1 + {n\over n+1} \sinh^2 \alpha)\qquad
 Q = cmn  \c\s.}
Of particular interest is
the extremal limit.  As before, the extremal limit corresponds to
$m \a 0, \ \alpha \a \infty $
such that $m\ss $ stays constant.
In the extremal limit, the horizon shrinks down to
zero size and becomes singular. The solution simplifies
to $$ ds^2 = \( 1+ {\tilde c M\over r^n}\)^{-1} (-dt^2 + dx^2)
     + dr^2 + r^2 d\Omega_{n+1} $$
$$    e^{-2\phi} = 1+ {\tilde c M \over r^n}$$
\eqn\extbkst{    B_{xt} = {\tilde c M\over r^n+ \tilde c M} }
where $\tilde c={n+1\over n} c$. Like the electrically charged black hole,
the transverse space is flat in the extremal limit. The extremal solution
has an extra symmetry which is not present in \bksth{}: It is boost
invariant in the $x,t$ plane.
Most importantly, \extbkst\ is precisely the solution found
by Dabholkar et.al.~\dghr\ describing the field outside of a fundamental
string.
So {\it a  straight fundamental string can be viewed as an extremal
black string}~\host.
This is not just a consequence of symmetry considerations.
One expects the solution for
a straight string to be static, translationally
invariant and  spherically symmetric. However, since it does not have a
regular horizon, it did not have to be contained in the family
of solutions\foot{The general static, spherically symmetric solution
to \simaction\ (without a horizon) is known in closed
form~\hohomass\ and contains one extra parameter than the black hole solutions
\phinz.}~\bksth{}.
Even given that it is contained in this family of solutions,
there is no a priori reason to expect that it would correspond to the extremal
limit.
Indeed, the analogous result is false in general relativity:
An electron cannot be viewed as an
extremal \RN\ black hole.

If $x$ is compact, the straight string can be viewed as an unexcited string
with
winding number one.
The fact that  this
string can be viewed as an extremal black string
strongly suggests that excited string winding states are black
strings and excited nonwinding states are black holes. In this regard, it is
important to keep in  mind the following simple observation. Given a particle
of mass $m$, there are two length scales that can be defined. One is the
Compton wavelength $ \lambda_{QM} = h/mc$ which can be thought of as
a quantum mechanical length scale. The second is the Schwarzschild radius
$ \lambda_G = Gm/c^2 $ which can be though of as a gravitational length
scale. When $\l_{QM} \gg \l_G$ i.e. when $m$ is much less than the Planck mass,
it is reasonable to ignore gravity and use quantum field theory in flat
spacetime as one usually does. But for $\l_G \gg \l_{QM} $ this is a terrible
approximation. All massive states in string theory satisfy this second
inequality. It
suggests that a better way to treat them might be to start with the black hole
solution and quantize about it.

In the dual picture, the extremal limit corresponds to boosting the
uncharged string to the speed of light. The resulting metric takes the form
\eqn\bobkst{  ds^2= -\(1-{\tilde cM \over r^n}\)dt^2
 + {2\tilde c M\over r^n} dt \,dx +\(1+{\tilde cM \over r^n}\)dx^2
 + dr^2 +r^2 d\Omega_{n+1}\>. }
 With new coordinates $x={1 \over 2} (u-v)$ and
 $ t = {1\over 2}(u+v)$, the metric becomes
\eqn\boobkst{  ds^2 = - du \,dv + dr^2+r^2 d\Omega_{n+1} +
	 {\tilde c M \over r^n} du^2.}
Metrics of this type are called plane
fronted waves. Like most solutions we have discussed so far, the extremal
black string~\extbkst\ is only a solution to the low energy field equations.
When higher powers of the curvature are included, it will have higher order
corrections. Similarly,
the duality transformation itself \sigmadual\ has higher order
corrections. Remarkably, these two corrections cancel each other!
It has been shown~\amkl\steif\ that~\boobkst\  is a solution to string theory
even including all higher order terms in the field equation.
The reason is essentially that the curvature is
null and so all powers of it vanish.
Since this represents the same conformal field theory (for compact $x$)
as \extbkst, one can also view
fundamental strings as strings moving at the speed of light.

It is known that
spacetime duality is accompanied by interchanging the momentum
and winding modes of test strings propagating in the background.
Since the extremal black string can be viewed as the field generated
by a pure winding state of the string, can its dual be interpreted as
the field generated by a pure momentum state? In other words,
does spacetime duality, in this case,  simply correspond to
interchanging the
momentum and winding of the source string? At first sight this interpretation
looks promising since the classical constraint equations for a string
show that for a pure
momentum state, the momentum must be null: An unexcited string always moves
at the speed of light. However an unexcited string is a pointlike object.
The field outside of a point particle accelerated to the speed of light is
given by~\boobkst\ with and extra $\delta(u)$ added to $g_{uu}$~\aise.
It does not have a spacelike translation symmetry and hence does not
have a spacetime dual. The solution~\boobkst\ describes an entire string
boosted to the speed of light, not a point particle.

It is interesting to consider the stability of black strings. Perturbations of
the simplest case, four dimensional Schwarzschild cross \R, have been studied
in detail~\grla. The conclusion is that certain modes with sufficiently
long wavelengths along the string grow exponentially with time. It appears that
the black string is trying to split into a series of separate disconnected
black holes. However this cannot happen since event horizons cannot bifurcate.
If we consider the black strings with charge, there is another reason why
they cannot split up. Recall that the charge is given by the integral of
$^*H$ over a $D-3$ sphere at large transverse distances. This charge is
conserved, but if the black string split into disjoint black holes, there
would be a nonsingular surface spanning the $D-3$ sphere, and the charge would
vanish by Stokes' theorem. At this time,  the significance of the unstable
modes is not yet clear. Are there stable black string solutions which are not
translationally invariant along the string?

For five dimensional black strings, the extremal limit
is shown in \tnaksing. This is the familiar spacetime of a naked singularity.
However, for $D>5$, the extremal spacetime resembles the extremal charged
black hole in four dimensions  \tnullsing.
The singularity splits into two parts, both of which are null.
Another difference is with the Hawking temperature.
One can compute a Hawking temperature of a black string by the usual
analytic continuation in time. In fact the formula derived earlier for
black holes \gentemp\
is still applicable since $g_{xx}$ remains finite and nonzero
at the event horizon. Applying this formula to \bksth{}\ one finds
\eqn\tempbkst{ T_H = {n \over 4\pi m^{1/n} \c}}
Since $m\a 0$ and $m\c^2$ stays constant in the extremal limit, we see that
for $n=1$ ($D=5$), the Hawking temperature of black strings diverges in this
limit, while for $n=2$ ($D=6$), it approaches a constant.
For $n>2$ ($D>6$) the situation is similar to \RN\ and the temperature
goes to zero. The fact that the temperature diverges in the extremal limit
for a five dimensional black string
is quite worrisome for it appears that Hawking
radiation will overshoot and end up with a naked singularity. But a similar
situation occurs for black holes with a different coupling between the
dilaton and the Maxwell field. The perturbations around these black holes
have been studied by Holzhey and Wilczek~\howi\ and they find that
large potential barriers form outside the black black holes which go to
infinity in the extremal limit. So even though the temperature is diverging,
 the energy radiated to infinity vanishes. If a similar thing happens
 here, the black string would only asymptotically reach its extremal limit.

It is interesting to note that it may actually
be easier to do a complete calculation of Hawking evaporation for
black strings than for ordinary black holes. This is because of two
factors. First,
like the electrically charged black holes, the coupling defined by the dilaton
is becoming weak at the horizon which may suppress
quantum effects in general.
Second, the solution is approaching the field outside of
a fundamental string and we know how to describe strings quantum mechanically.
A black string is likely to approach its extremal limit since the charge $\Q$
cannot be radiated away using point particles. This charge
will change only if one radiates
infinite strings.

More general black string solutions can be constructed along the lines of
Sec. 4. For example, black strings with rotation~\hohorot\
or electric and magnetic
charge~\senemst\ have been found. Solutions describing waves travelling
along an extremal black string have also been constructed \garfbkst.

\subsec{Four dimensions}

We now consider four dimensional black strings.  Unfortunately this
discussion will be very short since there aren't any.
To apply the construction in Sec. 2.3
one must start with a three dimensional black hole. But we saw in Sec. 4.6 that
there are no three dimensional solutions of low energy string theory
describing black holes.
Even if one relaxes the field
equations, one can show that there are no static  black
strings in any theory satisfying the dominant energy condition:
$T_{\mu\nu} t^\mu \t t^\nu \ge 0$ for all future directed timelike vectors
$t^\mu,  \t t^\nu$. This is an immediate consequence of a theorem due to
Hawking \hael\ which states that the event horizon of any
stationary black hole in such a theory
must be topologically $S^2$. If there was a static black string, one
could periodically identify to obtain a black hole with topology $T^2$.

This raises the following puzzle. As we have discussed, there is a two
dimensional black hole solution. One can always take its product with $T^2$ to
obtain a four dimensional solution with a toroidal event horizon. Why
doesn't this contradict Hawking's theorem?  This spacetime is not
asymptotically flat in the usual sense, but
Hawking's proof only involves a local calculation in the neighborhood of the
horizon. It should apply to spacetimes with this asymptotic behavior.
Since the dilaton is finite at the horizon, the Einstein metric
will also contain a toroidal horizon.
What about the dominant energy condition? Although
this condition is not satisfied in the string metric, it is
satisfied when the equations are reexpressed in terms of the Einstein
metric, and  the constant $\Lambda = 0$. But
 the two dimensional black hole only exists if $\Lambda$
is nonzero and positive. When reexpressed in terms of the Einstein metric,
this corresponds to a negative potential for $\phi$ which violates the
energy condition and allows the toroidal horizon.

\subsec{Three dimensions}

Finally we turn to three dimensions.  Black strings do exist in three
dimensions
if one
includes $\Lambda >0$ and
allows the dilaton to grow linearly at infinity. This is not surprising
since one can simply take the product of the two dimensional black hole
with $\R$. This yields a  black string without charge. To add charge, we
can simply follow the above example of boosting and then dualizing. The
result is \hhs\hoho
$$ ds^2 = -\(1-{M\over r}\) dt^2 + \(1-{\Q^2 \over Mr}\) dx^2 +
   {kdr^2 \over 4(r-M)
  (r-\Q^2/M)}$$
$$ e^{-2\phi} = r\sqrt k$$
\eqn\lbkst{ B_{xt} = \Q/r }
Setting $\Q=0$ we clearly  recover the two dimensional black hole cross $\R$.
Unlike the higher dimensional examples, the exact conformal field theory
is known~\hoho. Recall that
Witten showed that the exact conformal field theory associated with the
two dimensional black hole  could be described in terms of a gauged WZW
model in which one gauges a one dimensional subgroup of SL(2,\R). Similarly,
the exact CFT associated with the black strings can be obtained by starting
with the group $SL(2,\R) \times \R$ and gauging the same one dimensional
subgroup
of $SL(2,\R)$ together with a translation of $\R$.

This simple metric (and simple
construction) has a very rich global structure.
For $\Q<M$, the spacetime is similar to
the \RN\ solution.
There is an event horizon at $r=M$
and an inner horizon at $r=\Q^2/M$. The singularities are timelike and the
spacetime is timelike geodesically complete.  Note that the direction
along the string becomes timelike near the singularity.
There is also a region beyond the singularity corresponding to $r<0$
in \lbkst. In this case, the light cones
do not turn over on the other side of the singularity.
The WZW construction directly
gives you only two copies of the spacetime with identifications along the
inner horizon. But the universal covering space would have an
infinite number of copies.

Thus, like the rotating black hole of Sec. 4.3, the presence of the dilaton
does
not cause the inner horizon to become singular. Does the exact solution have
an inner horizon? It is not yet clear. On the one hand,
  (if $x$ is compact) this solution
 is supposed to  be equivalent to the boosted uncharged black string
which does not have an inner horizon. This is consistent with the
existence of an inner horizon in the low energy solution since this
horizon is unstable. Even though the curvature can be made small (by taking
$k$ large) the higher order corrections may become large. On the other hand,
the exact metric and dilaton (but not antisymmetric
tensor) for the three dimensional black string has been found \sfetsos\
and does have an inner horizon. Is it possible that an inner horizon
might not be well defined in string theory? Equivalent (exact)
solutions might differ
on whether there is an inner horizon or not.

\ifig\texthree{The extremal three dimensional black string has a horizon but
no singularity.}
{{\epsfysize = 5in \epsfbox{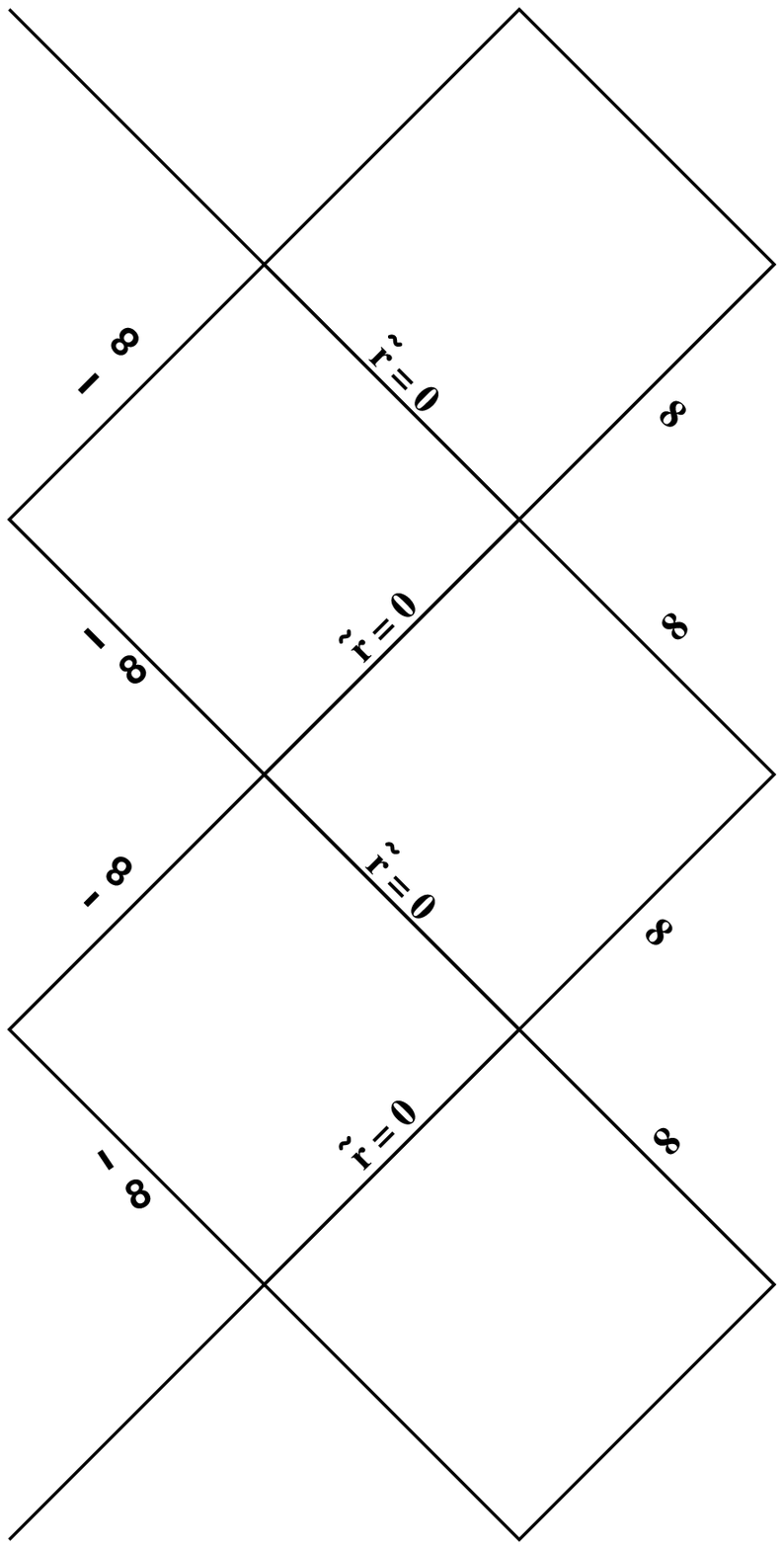}}}

For $\Q=M$, the horizons in \lbkst\ coincide. The metric becomes boost
invariant
as in the higher dimensional cases. It looks like the coordinates
$t$ and $x$ switch roles,
but this is misleading. One can show that geodesics
never reach $r<M$. The correct
extension across $r=M$ is in terms of a new radial
coordinate $\tilde r^2 = r-M$.
The resulting metric is
\eqn\metlimit{ds^2 ={\t r^2 \over \t r^2 +M} (-dt^2 + dx^2) +
    {k \, d\t r^2 \over  \t r^2 }}
This metric has the unusual property of having a horizon but no
singularity. (I know of no analog in general relativity.) This is just the
opposite of the extremal black strings in higher dimensions which had
a singularity but no horizon. The Penrose diagram looks like \texthree. The
regions on
both sides of the horizon are identical.
In the dual description, the extremal limit again corresponds to boosting
the uncharged string up to the speed of light.

Finally, for $\Q>M$, the metric appears to change signature at $r = \Q^2/M$.
But this is just another indication that an incorrect extension is being
used. The point $r=\Q^2/M$ turns out to be a conical singularity which can
be removed by making $x$ periodic.
The resulting space is completely nonsingular.
Surfaces of constant $t$ look like infinite
cigars. The exact conformal field theory associated with this is obtained
by gauging a different
subgroup of $SL(2,\R)$ together with translations in $\R$ where $\R$ is now
timelike \hoho.

\newsec{DISCUSSION}

To summarize,  we have investigated black hole and black string solutions
to low energy string theory. Both classes of solutions have
many unusual properties. For black holes, one of the
most important is that the string metric describing an extremal magnetically
charged black hole has neither a horizon nor a curvature singularity.
The spacelike surfaces contain infinite throats. These throats remain
when one adds electric as well as magnetic charge (provided that there is
only one Maxwell field) and when one adds a mass to the dilaton (provided
the black hole is sufficiently small). The infinite throats do not
remain when rotation is included. An  inner horizon is present only in certain
cases including nonzero rotation or large black holes with a massive
dilaton. Finally, extremal black holes (with charge of the same sign) have no
force between them when the dilaton is massless, but become repulsive
when the dilaton is massive.

Black string solutions are perhaps of less direct physical interest since,
as we have seen, they do not exist in four dimensions. But the fact that
in higher dimensions their extremal limit is equivalent
to an elementary  string, indicates that
by studying them, one might gain a deeper understanding of the fundamental
nature of strings. In three dimensions, their extremal limit corresponds to
a spacetime with the unexpected property of having an event horizon but
no singularity.

Since most of the solutions we have discussed only solve the low energy
equations of motion, one cannot use them to learn about singularities in
string theory. However, there are a few exceptions. As we have discussed,
the exact conformal field theory corresponding to the three dimensional
black string is known. When $\Q > M$, this solution is nonsingular.
Yet, applying the duality transformation
\sigmadual, one obtains a solution which
has a curvature singularity \hoho. (For a discussion of spacetime duality in
the context of gauged WZW models see \giveon.)
Since these are supposed to correspond to the
same conformal field theory, one is led to the conclusion that certain
curvature singularities do not  adversely affect string theory\foot{Since the
duality transformation is not exact there is a small possibility that the
exact dual solution will not have a curvature singularity.}! String
scattering in such a background is completely well defined (since it can be
calculated in the equivalent nonsingular spacetime). A simpler example of this
is to start with Minkowski space in cylindrical coordinates
$ds^2 = -dt^2 + dx^2 + dr^2 + r^2 d\theta^2$. Applying the duality
transformation to $\theta$ changes $r$ to $1/r$ and creates a curvature
singularity at $r=0$. Yet this solution is equivalent to flat spacetime.

However there are other examples of curvature
singularities which {\it do} affect strings. These are gravitational plane
waves with diverging amplitude. One can show that a string propagating through
such a wave becomes infinitely excited \steif\desa.
Physically, this is just a result of
the gravitational tidal forces. The singularity in the exact two dimensional
black hole is expected to be similar, but this has not yet been conclusively
demonstrated.

We conclude with some open problems:

1) It would be of great interest to
find the exact solution to string theory which approaches Schwarzschild
(or the charged black holes of Sec. 3) when the higher order corrections
to the field equations become unimportant. There have been several attempts
to use the gauged WZW approach to find such a solution \exactsol\basf.
Although new exact solutions have been obtained this way, so far none
can be interpreted as an asymptotically flat four dimensional black hole.

2) As we have just mentioned, there are two types of curvature singularities
in string theory. One affects strings and the other does not. What is
the essential difference between them? How generic are they in solutions
to string theory, and which type occurs in exact black hole solutions?
The standard singularity theorems of general relativity do not
apply to string theory. Is there an
analogous theorem which does apply?

3) It was suggested
in Sec. 5 that the massive states of a string should correspond
to black holes or black strings depending on whether their winding number
is zero or nonzero. Can this be rigorously established? If so, one could
view the decay of a massive string state as analogous to Hawking evaporation
(if the final states are massless) or quantum bifurcation of black holes
(if the final states are massive).

4) We have seen that many of the black hole solutions can be obtained by
solution
generating techniques. Although we have discussed only a few special cases,
for a spacetime with $d$ symmetry directions, the general transformation
(ignoring gauge fields)
is $O(d,d)$. However in general relativity, Geroch has shown \geroch\ that
there
is an infinite dimensional group which relates solutions to the vacuum Einstein
equations with two symmetry directions.  Can one similarly
extend the known $O(2,2)$ symmetry of low energy string equations
to an infinite dimensional group?

5) The full implications of spacetime duality have not yet been explored.
Strictly speaking we should talk only about duality invariant concepts.
For example, the three dimensional black string with $\Q<M$ indicates that
the inner horizon is not duality invariant. Is the event horizon duality
invariant? (If one applies the transformation \sigmadual\ to time
translations in Schwarzschild, one obtains another solution in which the
horizon becomes a singularity. But since the symmetry is not spacelike and
not compact, there is no proof that the two backgrounds correspond to the
same conformal field theory.)
There are also broader issues associated with the momentum-charge equivalence
\pqduality.
For example we usually think of momentum as associated with a spacetime
symmetry and $\Q$ as associated with an internal symmetry. The fact that they
are equivalent is a concrete indication of the unification of these symmetries
in string theory. As a second example, if $x$ is compact, $P_x$ should be
quantized. This implies $\Q$ should be quantized as well. This appears to be
a new argument for charge quantization.

6) Finally, there is the problem of calculating the Hawking evaporation of
black holes and black strings in string theory. This appears to be beyond
our current ability, although progress is being made.

\centerline{Acknowledgements}
It is a pleasure to thank my collaborators D. Garfinkle, J. Horne, A. Steif,
and A. Strominger. I have also benefited from discussions with I. Bars,
S. Giddings, R. Gregory,
J. Harvey, R. Kallosh, and C. Vafa. I thank the Aspen Center for Physics
where part of these lectures were written.
Finally, I wish to thank the organizers of the 1992 Trieste
Spring School for the invitation to lecture. This work was supported in
part by NSF Grant PHY-9008502.

\listrefs

\end